\documentclass[review,12pt,3p]{elsarticle} %

\usepackage{amssymb}
\usepackage{amsmath}
\usepackage{graphicx}
\usepackage{caption}
\usepackage{enumitem}
\usepackage{varwidth}
\usepackage{hyperref}
\usepackage{natbib}
\usepackage{subcaption}
\usepackage{appendix}
\usepackage{booktabs}
\usepackage{threeparttable}
\usepackage{float}
\usepackage{tcolorbox}
\usepackage{ragged2e}
\usepackage{url}
\usepackage[normalem]{ulem}
\usepackage{makecell}
\usepackage{xcolor}

\journal{Transportation Research Part D: Transport and Environment}

\begin{document}

\begin{frontmatter}



\title{Wildfire Evacuation Analysis Using Facebook Data: \\ Evidence from Palisades and Eaton Fires}

% Authors
\author[label1]{Shangkun Jiang}
\author[label2]{Ruggiero Lovreglio}
\author[label3]{Thomas J. Cova}
\author[label1]{Sangung Park}
\author[label4]{Susu Xu}
\author[label1]{Xilei Zhao\corref{cor1}}

% Corresponding author
\cortext[cor1]{Corresponding author: Xilei Zhao. Email: xilei.zhao@essie.ufl.edu}

% Affiliations
\address[label1]{Department of Civil and Coastal Engineering, University of Florida, Gainesville, FL 32611, US}

\address[label2]{School of Built Environment, Massey University, Auckland, 0632, New Zealand}

\address[label3]{School of Environment, Society \& Sustainability, University of Utah, Salt Lake City, UT 84112, US}

\address[label4]{Department of Civil and Systems Engineering, Johns Hopkins University, Baltimore, MD 21218, US}

%% Abstract
\begin{abstract}
The growing frequency and intensity of wildfires pose serious threats to communities in wildland-urban interface regions. Understanding evacuation behavior is critical for effective emergency planning. This study analyzes evacuation during the 2025 Palisades and Eaton Fires using high-resolution Facebook data. We propose a comprehensive framework to derive wildfire evacuation-related metrics, including compliance rate, departure timing, delay, origin-destination flows, travel distance, and destination types. A new metric, Damage-Evacuation Disparity Index (DEDI), identifies areas with severe structural damage but low evacuation compliance. Results reveal spatiotemporal heterogeneity: residents closer to the fire evacuated earlier, whereas late or nighttime orders led to lower compliance and longer delays. Contrasting patterns between East and West Altadena further illustrate this disparity. DEDI-identified communities exhibited higher social vulnerability and fire risk. Most evacuations concluded in residential areas, while longer trips concentrated in hotels and public facilities. These findings showcase the Facebook data's potential for data-driven wildfire evacuation planning.

\end{abstract}

% \begin{highlights}
% \item Facebook datasets reveal significant spatial and temporal heterogeneity in wildfire evacuation compliance rate.
% \item Residential areas dominate evacuation destinations, while hotels are associated with longer travel distance.
% \item Evacuation compliance rates exhibit a strong negative correlation with distance from the fire perimeter.
% \item Socioeconomic factors, including household income and racial composition, significantly influence evacuation compliance.
% \item The study proposes a transferable analytical framework for real-time evacuation research across disaster contexts.
% \end{highlights}

%% Keywords

\end{frontmatter}

\section{Introduction} \label{Introduction}
% Part I - Research problem:
Increased warming has led to hotter and drier weather conditions that, combined with urban expansion in the wildland-urban interface (WUI) area, have intensified wildfire frequency and expanded community vulnerability worldwide \citep{bowman2023taming, jones2024global, radeloff2018rapid, wu2022wildfire}. California exemplifies this crisis, with 2,857 large wildfires from 1950 to 2020 burning approximately 20.67\% of the state's land area \citep{calfire, ostertag2023investigating}. Most strikingly, the January 2025 Eaton and Palisades Fires in Los Angeles killed 29 people, destroyed over 15,000 structures, and displaced more than 130,000 people \citep{calfire}. These catastrophic impacts underscore the urgent need to advance wildfire evacuation research with a focus on WUI areas. Understanding human behavior during evacuations is essential for transportation planning, land use policy, and wildfire emergency management. In particular, analyzing human mobility patterns and spatial dynamics provides critical insights into evacuation compliance, destination choice, and travel distance. These factors directly affect the efficiency of evacuation operations and the resilience of affected communities. However, systematic evidence on these mobility characteristics remains scarce, limiting the ability to design effective evacuation strategies \citep{cova2024destination, zhao2022estimating}.

% Part II - What are the research gaps/challenges?
In recent years, the Meta AI for Good Program\footnote {Please refer to \url{https://dataforgood.facebook.com/} (accessed 2025.10.26).} \citep{maas2019facebook} has emerged as a valuable data source for disaster research \citep{maas2019facebook, jia2020patterns}. These datasets, hereafter referred to as Facebook data, are open and privacy-protected, offering population and mobility information for crisis-affected regions \citep{maas2019facebook, rashid2025understanding}. They capture evacuation dynamics during wildfires, including origin-destination (OD) and movement volumes \citep{jia2020patterns}. These features make Facebook data well-suited for systematic analysis of human mobility and spatial dynamics during wildfire. Facebook data has been successfully used to study human behavior during hurricanes \citep{jamal2023understanding, rashid2025understanding}, earthquakes \citep{varol2024movement}, and the COVID-19 pandemic \citep{bonaccorsi2020economic, galeazzi2021human}. Despite growing use of Facebook data in disaster research, no integrated framework has been developed to systematically derive evacuation-related metrics that capture people's choices of whether, when, and where to evacuate; this research gap is particularly evident for wildfire evacuations.

Another gap concerns community vulnerability to wildfires. Most existing studies on wildfire vulnerability treat structural vulnerability and social vulnerability as separate issues \citep{mahmoud2024leveraging, naser2025vulnerability, sun2024social}. Few have examined how these two dimensions jointly shape the overall community vulnerability to wildfires. As a result, vulnerable communities that experience severe structural damage but have low evacuation compliance remain poorly identified \citep{lambrou2023social}. These communities face even higher wildfire fatalities and injuries due to delayed or incomplete evacuation, often caused by insufficient warnings, limited transportation access, or socioeconomic barriers \citep{paveglio2016evaluating}. Developing a composite metric that integrates structural damage severity and evacuation compliance can help better capture these spatial disparities and identify communities in need of targeted support \citep{palaiologou2019social}.

The third gap lies in the limited understanding of the influence of land-use on wildfire evacuation travel distance, flow patterns, and destination choices during wildfires. While previous studies have examined evacuation decisions and timing \citep{forrister2024analyzing, grajdura2022fast, lovreglio2019modelling, mclennan2019should, sun2024social}, few have comprehensively analyzed how and where people move once they evacuate. Recent research has begun to explore evacuation routes \citep{brachman2020wayfinding, wong2023understanding} and destination choices \citep{cova2024destination}; however, comprehensive analyses that integrate mobility data with spatiotemporal evacuation patterns across diverse land-use contexts remain scarce. Understanding how land use shapes evacuation behavior is critical for emergency planning and shelter allocation \citep{cova2024destination, ronchi2019open, toledo2018analysis, kuligowski2021evacuation}.

% Part III - What did we do to solve this problem?
This study advances wildfire evacuation research by using Facebook data to analyze two major wildfires in Los Angeles (LA) in January 2025, including the Palisades and Eaton Fires. These events were chosen because they were the most destructive among the fourteen wildfires that took place that month in LA and ranked among the deadliest in Southern California's history. We derive multiple evacuation metrics, including evacuation rates, departure curves, delays, OD flows, destination distributions, and travel distance. In addition, we introduce a new metric, the Damage-Evacuation Disparity Index (DEDI), which integrates evacuation compliance with structural damage severity to identify highly vulnerable areas. The analysis is conducted at the Bing-tile level (approximately 2.4 km $\times$ 2.4 km), comparable to the size of a typical traffic analysis zone (TAZ) in Southern California \citep{SCAG2024ConnectSoCal}. Specifically, this study aims to address the following research questions:

\begin{itemize}
    \item How to develop a comprehensive framework that uses Facebook data to dynamically measure wildfire evacuations?
    \item Can community vulnerability be measured by integrating wildfire damage severity and Facebook data-derived evacuation compliance?
    \item How are land-use characteristics associated with evacuation travel distance, flow patterns, and destination choices during wildfire evacuations?
\end{itemize}

% Part IV - What is the structure of this paper?
The remainder of this paper is organized as follows: Section \ref{Literature review} reviews relevant prior research. Section \ref{Methodology} describes the analytical framework and methods employed. Section \ref{Data} introduces the case study of the Los Angeles wildfire in early January 2025 and outlines the Facebook data used. Section \ref{Results} presents the findings from population evacuation behavior and model results during wildfires. Finally, Section \ref{Discussion} and \ref{Conclusion} discuss key findings, strengths, limitations, and potential directions for future research.

\section{Literature review} \label{Literature review}
Wildfire evacuation research has evolved in its methodological approaches over the past decades. Early studies on wildfire evacuation relied on survey \citep{grajdura2021awareness, kuligowski2022modeling, toledo2018analysis, whittaker2017experiences, xu2023predicting} and interviews \citep{christianson2019wildfire, wong2020can, grajdura2021awareness} to understand evacuation decisions \citep{xu2023predicting, mccaffrey2018should}, timing \citep{mccaffrey2018should}, risk perceptions \citep{mccaffrey2018should, toledo2018analysis} and affected factors \citep{lindell2012protective, kuligowski2013predicting}. While these approaches provided valuable behavioral insights, they are constrained by recall biases, self-reporting limitations, high collection costs, and limited spatiotemporal resolution \citep{wu2022wildfire, zhao2022estimating}.

The emergence of mobile device location data provided an alternative analytical perspective for granular spatiotemporal analyses of evacuation behavior \cite{cova2024destination, raei2025data, wu2022wildfire, zhao2022estimating, sun2024social}. For example, \cite{wu2022wildfire} analyzed household evacuation decisions during the 2019 Kincade Fire using GPS data, extending the survey-based findings of \cite{kuligowski2022modeling} and illustrating how different data sources can offer complementary perspectives on evacuation behavior. Recent work has explored evacuation routes \citep{raei2025data} and destination choices \citep{cova2024destination}. However, these data sources involve evacuation behavior inference algorithms that may introduce systematic biases \citep{chen2014traces, zhao2022estimating, guan2025using}, remain costly to acquire, and are often inaccessible for broader research and operational applications \citep{liu2025hurricane}. Moreover, traffic detectors \citep{dixit2014evacuation, feng2022modeling, rohaert2023traffic, rohaert2023analysis} and connected vehicles data \citep{ahmad2024evaluating} have also been used to capture evacuation dynamics through network performance metrics, but provided limited spatial coverage, particularly in rural wildfire-prone regions \citep{liu2024association, zhao2022estimating}.

Due to the accessibility and coverage limitations of the above-mentioned data sources, researchers have resorted to Facebook data as a promising solution \citep{maas2019facebook, jamal2023understanding, jia2020patterns, rashid2025understanding}. As shown in Table \ref{tab: FB_data_app_review}, the Facebook population and movement datasets have been successfully applied across various disaster contexts, including earthquakes \citep{varol2024movement}, hurricanes \citep{jamal2023understanding, rashid2025understanding, rashid2025network}, and pandemics \citep{bonaccorsi2020economic, galeazzi2021human, beria2021presence, lusseau2023disparities}, supporting analyses of population displacement, movement patterns, and network dynamics. Furthermore, some prior works have shown sufficient representativeness of Facebook data across demographic groups and geographic regions \citep{jia2020patterns, duan2024identifying, rashid2025understanding}.

\begin{table}[!ht]
\centering
\caption{Applications of Facebook population and movement datasets across disasters and metrics used.}
\label{tab: FB_data_app_review}
\resizebox{\textwidth}{!}{%
\begin{tabular}{lllll}
\hline
\textbf{Literature} & \textbf{Disaster} & \textbf{Topic} & \textbf{Dataset} & \textbf{Metric} \\ \hline
\citet{jia2020patterns} & Wildfire & Population displacement & Population & \begin{tabular}[c]{@{}l@{}}1. Population count\\      2. Z-score of population\end{tabular} \\ \hline
\citet{varol2024movement} & Earthquake & Movement pattern & Movement & \begin{tabular}[c]{@{}l@{}}1. Movement count percent change\\      2. Travel distance\end{tabular} \\ \hline
\citet{jamal2023understanding} & Hurricane & Community resilience & Population & Population activity rate \\ \hline
\citet{rashid2025understanding} & Hurricane & Evacuation modeling & Population & Stay index \\ \hline
\citet{rashid2025network} & Hurricane & Traffic volume prediction & Movement & Movement percentage change \\ \hline
\citet{bonaccorsi2020economic} \\ \citet{galeazzi2021human} & COVID-19 & Mobility network analysis & Movement & \begin{tabular}[c]{@{}l@{}}1. Weakly connected components\\      2. Largest weakly connected component\end{tabular} \\ \hline
\citet{beria2021presence} \\ \citet{lusseau2023disparities} & COVID-19 & Population and mobility dynamics & Movement & \begin{tabular}[c]{@{}l@{}}1. Movement count percentage change\\      2. Travel distance\\      3. Z-score of population\end{tabular} \\ \hline
\citet{liu2024association} & /\ & Air quality and human mobility & Movement & Movement count percent change \\ \hline
\citet{duan2024identifying} & /\ & Counter-urbanisation patterns & Population & \begin{tabular}[c]{@{}l@{}}1. Population count\\      2. Population count percent change\end{tabular} \\ \hline
\end{tabular}%
}
\end{table}

Despite these applications in other disaster contexts, Facebook data remain underutilized in wildfire research. The only relevant study \citep{jia2020patterns} analyzed population displacement during California wildfires using built-in Z-score variables but did not examine evacuation-specific metrics such as departure timing, delay, travel distance, or destination patterns. Consequently, there remains a need for a comprehensive analytical framework that leverages Facebook data to derive evacuation-related metrics and evaluate evacuation behavior and mobility patterns during wildfires. Moreover, existing wildfire studies have rarely investigated the relationship between evacuation rates and structural damage \citep{mahmoud2024leveraging, palaiologou2019social}, which could reveal vulnerable communities that experience high damage yet low evacuation compliance due to socioeconomic or accessibility barriers \citep{lambrou2023social, mahmoud2020barriers}. Finally, little is known about how land-use characteristics influence evacuation mobility, such as travel distance, flow patterns, and destination choices, particularly when analyzed through Facebook data. While recent studies have examined evacuation routes \citep{brachman2020wayfinding, wong2023understanding} and destinations \citep{cova2024destination}, comprehensive analyses integrating evacuation OD flows with land-use variations and destination types are rare. This study aims to address these gaps by using open-access Facebook data to analyze evacuation mobility during the 2025 Palisades and Eaton Fires, offering a scalable approach to understanding disaster mobility in wildfire scenarios.

\section{Methodology} \label{Methodology}
\begin{figure}[!ht]
  \centering
  \includegraphics[width=\textwidth]{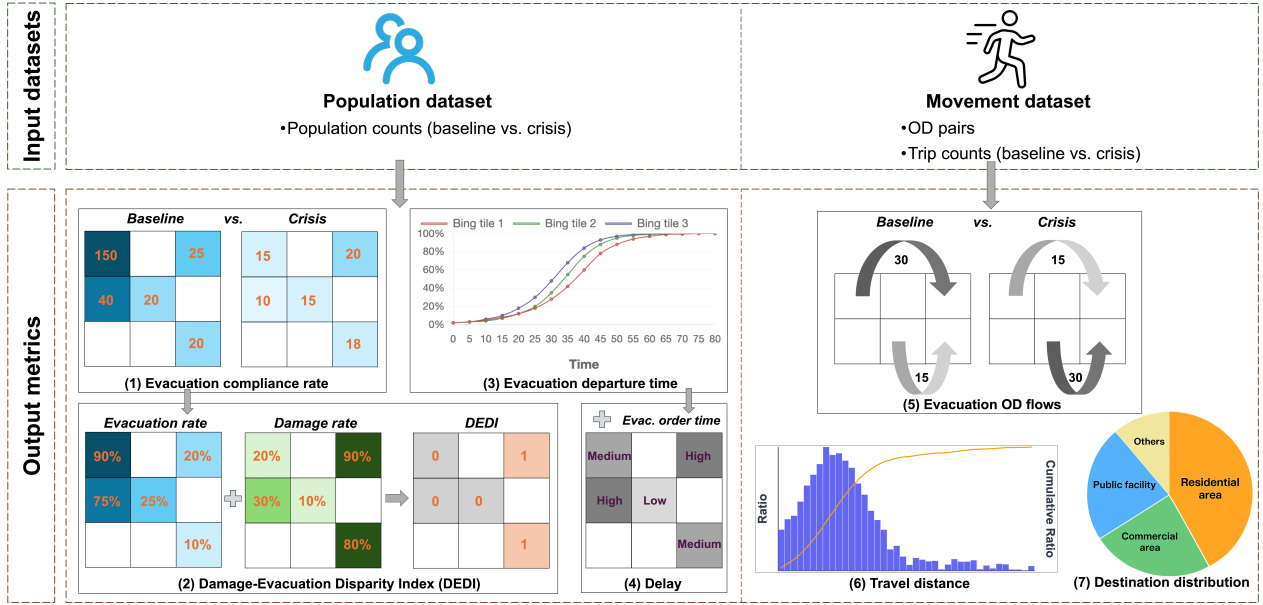}
  \caption{Proposed framework to analyze evacuation behavior using Facebook data.}
  \label{fig: Analysis framework}
\end{figure}

We propose a comprehensive framework for computing various wildfire evacuation metrics from Facebook data, as shown in Figure \ref{fig: Analysis framework}. The framework relies on two Facebook datasets. The first is the population dataset, which provides baseline and crisis population counts for each Bing tile. A Bing tile represents the spatial aggregation unit in the Facebook dataset, covering an area of approximately 2.4 km by 2.4 km. This spatial resolution is comparable to the typical TAZ used in Southern California \citep{SCAG2024ConnectSoCal}. The second is the movement dataset, which records origin–destination (OD) pairs and trip counts. Detailed descriptions of both datasets are provided in subsection \ref{Data description}. Using the population dataset, we estimate evacuation compliance rates by comparing baseline and crisis population counts for each Bing tile. We also combine evacuation rates with CAL FIRE damage inspection data\footnote{Please refer to \url{https://hub-calfire-forestry.hub.arcgis.com/datasets/cal-fire-damage-inspection-dins-data/explore} (accessed 2025.10.26).} to identify areas with high damage but low evacuation. These areas are captured through a binary indicator, the Damage-Evacuation Disparity Index (DEDI). In addition, temporal changes in population decline allow us to measure evacuation departure times and delays. Using the movement dataset, we derive evacuation OD flows by contrasting trip counts from baseline and crisis periods, calculate travel distance to assess spatial displacement patterns, and examine destination distributions by classifying trip-end Bing tiles. Together, these seven metrics provide a systematic view of wildfire evacuation behavior using Facebook datasets.

\subsection{Evacuation compliance rate}
The Facebook population dataset is employed to estimate the population evacuation compliance rate during wildfires. The population percent change is calculated between the baseline and wildfire periods using Equation \ref{eq: Population percentage change}. We assume that negative deviations from the baseline reflect evacuation behavior, while subsequent increases represent reentry as residents return \citep{jia2020patterns, rashid2025understanding}. A negative $P_{i}\left ( t \right )< 0$ indicates evacuation, as the population of Bing tile $i$ on day $t$ is below the baseline. A positive value of $P_{i}\left ( t \right )>0$ indicates an influx of population.

\begin{equation}
P_{i}\left ( t \right )=\frac{n_{i}^{crisis}\left ( t \right )  -n_{i}^{baseline}\left ( t \right ) }{n_{i}^{baseline}\left ( t \right ) }
\label{eq: Population percentage change}
\end{equation}

\noindent Where $P_{i}\left ( t \right )< 0$ is the population percentage change for Bing tile $i$ at day $t$, and it ranges from $\left[-1,+\infty\right)$.

Based on these percent changes, we define the evacuation compliance rate, $E_{i}\left ( t \right )$, for Bing tile $i$ at time $t$ as the absolute share of Facebook users who left the tile. This metric builds on Equation \ref{eq: Population percentage change} but considers only negative changes. We use absolute values to ensure the evacuation compliance rate is non-negative. While $E_{i}\left ( t \right )$ can be tracked for every Bing tile to illustrate the temporal dynamics of out-migration, the value should be represented as a single, fire-specific scalar that captures residents' compliance with evacuation orders. To achieve this, we define a reference time $t^{*}$, as the timestamp when the order produces the maximum population deficit. The evacuation compliance rate for tile $i$ is then computed as $E_{i}\left ( t^{*} \right )$. Guided by the issuance timeline in Table \ref{tab: timelineofwildfire}, we set $t^{*}$ for both wildfires to January 8, 2025 at the time slice from 8 AM to 4 PM. Using $E_{i}\left ( t \right )$ in this way ensures that the reported evacuation compliance rate reflects actual adherence to the evacuation order rather than daily fluctuations in population counts.

\begin{equation}
E_{i}\left ( t \right ) =\begin{cases}
\left | P_{i}\left ( t \right ) \right | ,  & \text{ if } n_{i}^{crisis}\left ( t \right ) < n_{i}^{baseline}\left ( t \right )  \\
0,  & \text{otherwise}
\end{cases}
\label{eq: evacuation compliance rate}
\end{equation}

We next identify wildfire vulnerability hotspots by combining structure damage levels with evacuation compliance rates for each Bing tile. Damage data are obtained from the CAL FIRE Damage Inspection (DINS) dataset, which records the severity of structure damage near fire perimeters. The dataset categorizes damage into five classes: Destroyed (>50\%), Major (26–50\%), Minor (10–25\%), Affected (1–9\%), and No Damage. For each Bing tile (2.4 km $\times$ 2.4 km), we compute the average damage level of all contained structures and compare it with the corresponding evacuation compliance rate. To capture tiles where residents faced severe property loss but exhibited low evacuation responses, we construct the Damage-Evacuation Disparity Index (DEDI), as defined in Equation \ref{eq: DEDI}. A Bing tile is flagged as a wildfire vulnerability hotspot when it records both high structural damage and low evacuation compliance, indicating a mismatch between objective risk and evacuation behavior.

\begin{equation}
DEDI_{i} =\begin{cases}
1,  & \text{ if } D_{i}>E_{i}\left ( t^{*} \right )\\
0,  & \text{ otherwise } 
\end{cases}
\label{eq: DEDI}
\end{equation}

\noindent Where $D_{i}\left ( t \right )$ denote the damage rate for area $i$.

\subsection{Evacuation departure time and delay}
To examine evacuation departure patterns during wildfire events, we define the departure population $D_{i}\left ( t \right )$ for each Bing tile $i$ at time $t$ as the difference between the baseline and crisis populations (Equation \ref{eq: departure population}). The cumulative departure percentage $C_{i}\left ( t \right )$ for tile $i$ up to time $t$ is then calculated as the proportion of the cumulative sum of departures by time $t$ relative to the total departure population of that tile (Equation \ref{eq: cumulative departure percentage}). These functions produce evacuation curves that describe the temporal dynamics of population departures in response to evacuation orders at the grid-cell level.

\begin{equation}
D_{i}\left ( t \right )=max\left ( n_{i}^{baseline}\left ( t \right ) -n_{i}^{crisis}\left ( t \right ),0 \right ) 
\label{eq: departure population}
\end{equation}

\begin{equation}
C_{i}\left ( t \right )= \frac{ {\textstyle \sum_{\tau \le t}^{}} D_{i}\left ( \tau  \right )}{ {\textstyle \sum_{\tau}^{}} D_{i}\left ( \tau  \right )} \times 100\%
\label{eq: cumulative departure percentage}
\end{equation}

We also define a delay metric for each Bing tile to quantify evacuation delay. The time when the evacuation order reaches tile $i$ is denoted as $t_{r} $. At each time slice $t$, the delay is computed as the product of the departing population and the elapsed time since $t_{r} $. Given an analysis end time $T_{e}$, the total delay for tile $i$ is calculated as Equation \ref{eq: delay}.

\begin{equation}
T_{i}=  \sum_{t=t_{r}}^{T_{e}}\left ( D_{i}\left ( t \right )\times max\left ( t- t_{r},0 \right ) \right ) 
\label{eq: delay}
\end{equation}

\subsection{Evacuation OD flows and destinations} \label{section: evac_mov}
\begin{figure}[!ht]
  \centering
  \includegraphics[width=0.9\textwidth]{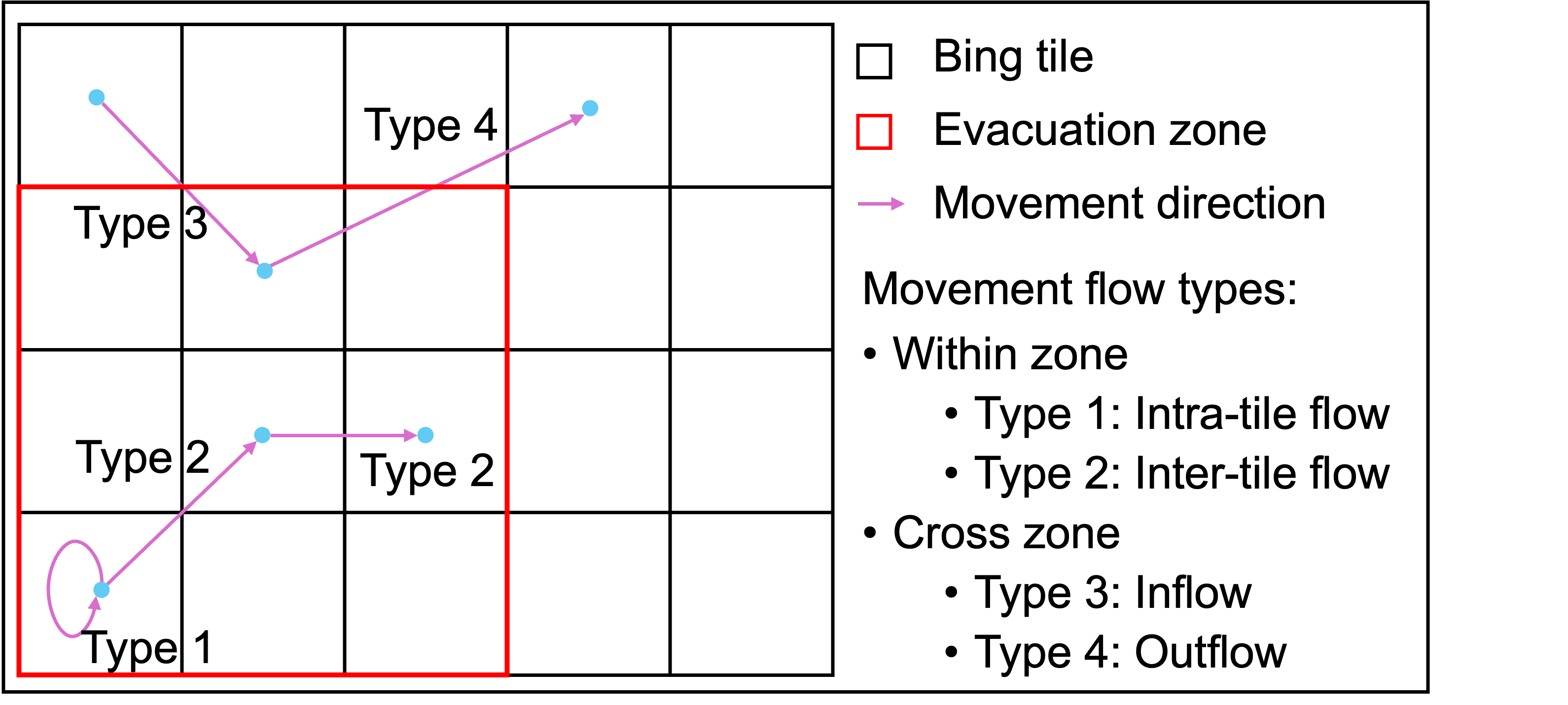}
  \caption{Movement types classification using Bing tiles (2.4 km $\times$ 2.4 km).}
  \label{fig: Different_movement_types}
\end{figure}

The Facebook Movement dataset is used to analyze evacuation movement behavior during wildfires. The analysis focuses on evacuation OD flows, travel distance, and destination distributions. Movements are classified based on their spatial relationship to evacuation zones, as shown in Figure \ref{fig: Different_movement_types}. This classification distinguishes movement types and identifies evacuation-related flows. Evacuation movements are defined as OD flows that originate in evacuation or warning zones and terminate in external areas (Type 4 in Figure \ref{fig: Different_movement_types}). They also include inter-zone movements where destinations are located farther from the wildfire perimeter than origins during the evacuation period. The wildfire evacuation period is defined as the three days following the issuance of an evacuation order, which is the typical time window for wildfire evacuation \citep{cova2024destination, jia2020patterns, zhao2022estimating}.

After identifying all evacuation movements, we examine their characteristics, including percent change compared with baseline, travel distance, and destination distribution. Evacuation movements are grouped by OD pairs to generate OD flows. We then calculate the percent change in the number of these flows during the wildfire period relative to the baseline and examine the relationship between evacuation movement counts and distance. This approach allows us to examine how wildfire events influence evacuation movement behavior.

Based on evacuation movements, we then investigate the types of evacuation destinations and their relationship with evacuation distance. Destinations are classified using parcel-level land-use data, following the classification criteria in Table \ref{tab: landClass} \citep{cova2024destination}. For each destination Bing tile, we calculate the proportion of land-use types according to their area coverage. For each OD pair $ o\in \mathcal{O}$, we multiply the total number of evacuation trips $w_{o}$ by the land-use proportion ${\lambda} _{o}^{k}$ of type $k$ in the destination tile. Summing across all OD pairs yields the overall share of evacuation movements to land-use type $k$, denoted as $S_{k}$ (Equation \ref{eq: destination_share}).

\begin{equation}
S_{k} = \frac{\sum_{o \in \mathcal{O}} w_{o} \cdot \lambda_{o}^{k}}{\sum_{o \in \mathcal{O}} w_{o}}
\label{eq: destination_share}
\end{equation}

\noindent Where $\mathcal{O}$ is the set of all OD pairs, $w_{o}$ is the number of evacuation trips associated with OD pair $o$, and $\lambda _{o}^{k}$ is the proportion of land-use type $k$ within the destination tile of OD pair $o$.

To compute the mean evacuation distance for each land-use type $k$, we selected all OD pairs whose destination contains land-use type $k$, denoted as $\mathcal{O}_{k} \subseteq  \mathcal{O}$. The weighted average evacuation distance are then given by Equation \ref{eq: mean distance}. Finally, we conducted Pearson's correlation analysis to examine the relationship between evacuation distance and the type of destination land use.

\begin{equation}
\bar{d} _{k} =\frac{ {\textstyle \sum_{o\in {\mathcal{O}_{k}}}^{}}w_{o}  \cdot d _{o}}{ {\textstyle \sum_{o\in \mathcal{O}_{k}}^{}} w_{o}}
\label{eq: mean distance}
\end{equation}

\noindent Where $d_{o}$ is the travel distance for OD pair $o$.

\section{Data} \label{Data}
\subsection{Study site} \label{Study site}
% Southern California wildfires and research area
In January 2025, a series of wildfires struck Southern California, severely impacting the Los Angeles metropolitan area and nearby counties \citep{calfire}. Fourteen fires ignited under conditions of drought, low humidity, and hurricane-force Santa Ana winds \citep{berker2025}. By the end of the month, the fires had burned more than 57,000 acres, caused at least 29 fatalities, displaced over 200,000 residents, and destroyed more than 18,000 homes and structures \citep{calfire, berker2025}. Among them, the Eaton fire in Altadena and the Palisades fire in Pacific Palisades were the most destructive. Both started on January 7 and continued for 24 days \citep{calfire}. Table \ref{tab: timelineofwildfire} summarizes the key dates and impacts.

The Palisades fire was reported around 10:30 PST near the Pacific Palisades neighborhood of Los Angeles. Evacuation orders were issued ten hours later and eventually extended to the Pacific Coast Highway, Tarzana, and Encino, displacing more than 30,000 residents. The fire burned 23,707 acres, destroyed 6,837 structures, and caused 12 deaths. It ranked as the tenth deadliest and third most destructive wildfire in California, and the most severe in Los Angeles history \citep{Caramela2025}. Later that evening at 18:18, the Eaton Fire ignited in Altadena (north of Pasadena), about 40 miles away. Evacuation orders were issued six hours later, affecting Pasadena, northern Sierra Madre, and Arcadia. More than 100,000 residents were evacuated within 24 hours. Because the fire occurred at night and the community west of Lake Avenue in Altadena did not receive an evacuation order until 3:25 am. (more than nine hours after ignition) \citep{washpost2025altadena, yahoo2025altadena}, all 17 fatalities from the Eaton fire occurred west of Lake Avenue \citep{nbc2025eaton}, with 15 of them in the area that received its first evacuation order at 3:25 a.m. The fire was fully contained on January 31, 2025, after burning 14,021 acres, destroying 10,491 structures. It ranks among the most destructive wildfires in Southern California's history.

\begin{figure}[!ht]
  \centering
  \includegraphics[width=\textwidth]{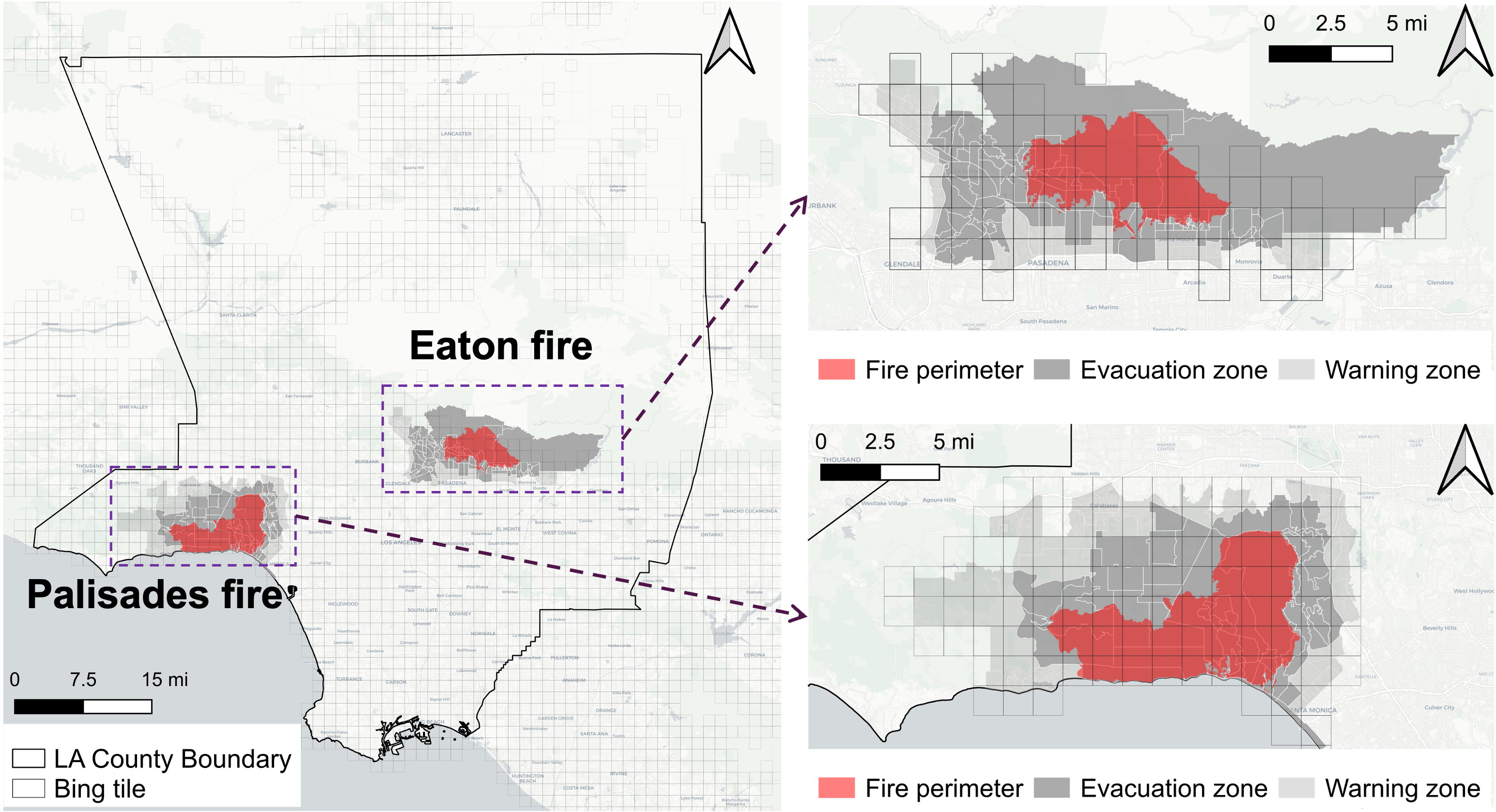}
  \caption{Perimeters of the Eaton and Palisade fires and their warning and evacuation zones in the LA County. \textit{Note}: the Bing tile dimension is 2.4 km $\times$ 2.4 km.}
  \label{fig: Research_area}
\end{figure}

This study focuses on the Palisades and Eaton fires and their corresponding official evacuation/warning zones. Figure \ref{fig: Research_area} shows the study area and the Bing tiles used for analysis. Figure \ref{fig: evac_order_distribution}(a) presents the timing of the evacuation order issuance and its corresponding spatial coverage, and Figure \ref{fig: evac_order_distribution}(b) shows the distribution of order issuance times for each Bing tile. The data are described in detail in Section \ref{Data description}.

\subsection{Facebook data description} \label{Data description}
\begin{figure}[!ht]
  \centering
  \includegraphics[width=\textwidth]{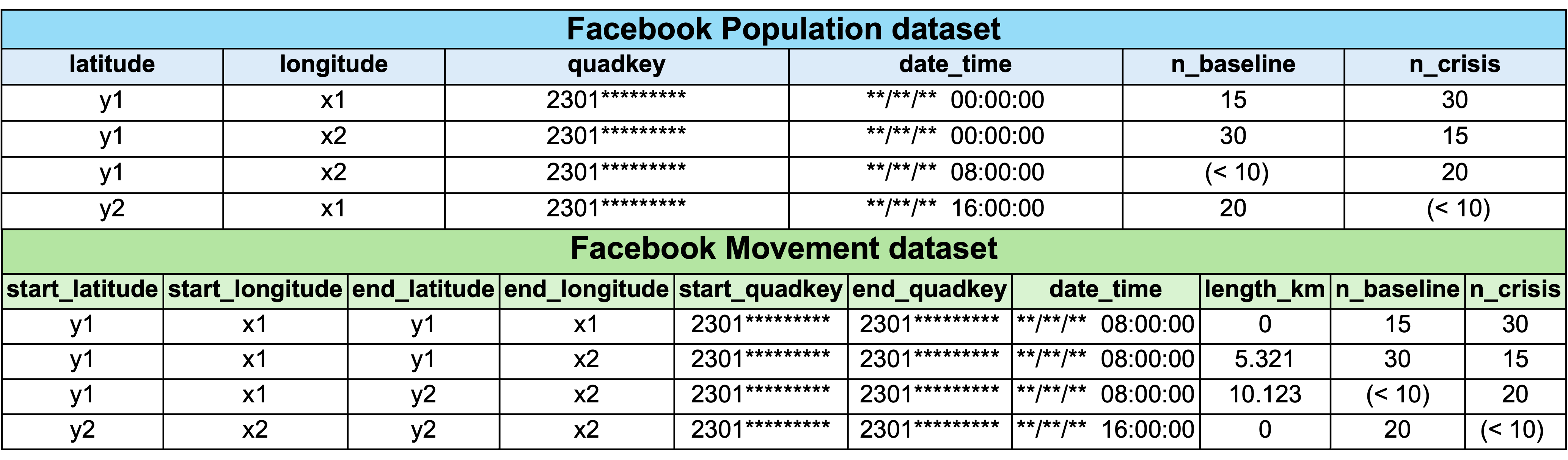}
  \caption{Facebook population and movement data structure.}
  \label{fig: Data_structure}
\end{figure}

% What are the datasets we use?
This study uses the Facebook population and movement datasets for analysis. The observation window extends from January 7 to March 10, 2025 and covers the critical dates listed in Table \ref{tab: timelineofwildfire} \citep{calfire}. The datasets cover most areas in California where Facebook APP's location signals are available, as shown in Figure \ref{fig: Research_area}. As shown in the data structure of Figure \ref{fig: Data_structure}, both datasets are temporally and spatially aggregated. Counts are summarized in eight-hour intervals and mapped to level-14 Bing tiles, which represent grid cells of roughly 2.4 km $\times$ 2.4 km \citep{bingMapsTileSystem, maas2019facebook}. The population dataset reports the number of users with location services enabled in each tile during every interval, whereas the movement dataset records the number of movements between pairs of tiles; each tile is identified by a unique \textit{quadkey}. For each Bing tile, \textit{n\_baseline} gives the mean population or movement count during a 45-day pre-event period \citep{maas2019facebook}. Facebook also derives three additional metrics: the absolute difference between crisis and baseline counts (\textit{n\_difference}), the percentage change (\textit{percent\_change}), and a z-score that standardizes deviations from baseline levels \citep{maas2019facebook}.

% What are the pros of these datasets? - privacy protection mechanism
Both datasets include several privacy safeguards that extend beyond simple data aggregation \citep{maas2019facebook}. First, a row is removed if both the baseline ($n\_baseline$) and crisis ($n\_crisis$) counts are below 10. If only one count meets or exceeds this threshold, that value is retained, and the other is set to null, together with any derived metric. Second, random noise is added to the crisis counts to prevent re-identification in sparsely populated areas. Third, the data are spatially smoothed with inverse-distance weighting \citep{achilleos2011inverse, maas2019facebook}, which blends each location's count with those of nearby tiles, giving greater weight to closer neighbors. These data mechanisms protect users' privacy but also add uncertainty to the estimates. Despite this drawback, Facebook datasets provide valuable insights for research on human behavior during disasters \citep{jia2020patterns, jamal2023understanding, varol2024movement, rashid2025understanding}.

Missing values in the Facebook dataset resulted from privacy protection mechanisms that required imputation. For each Bing tile, we calculated the proportion of missing values for both \textit{n\_baseline} and \textit{n\_crisis}. Tiles (\textit{quadkey}) with more than 14\% missing observations (equivalent to over one day per week) were excluded from the analysis. We assumed that counts in adjacent 8-hour windows exhibited minimal variation and temporal continuity \citep{maas2019facebook, duan2024identifying}. Therefore, each missing value was imputed using the moving average method, calculated as the mean of the corresponding time interval on the previous and following days \citep{yabe2020effects, jia2020patterns}. The panel datasets were then used for subsequent analysis.

\section{Results} \label{Results}
This section presents the empirical results of the evacuation metrics defined in Section \ref{Methodology}, including evacuation compliance rates, departure times, delays, evacuation OD flows, travel distance, and destination types.

\subsection{Evacuation compliance rate}
\begin{figure}[!ht]
  \centering
  \includegraphics[width=0.8\textwidth]{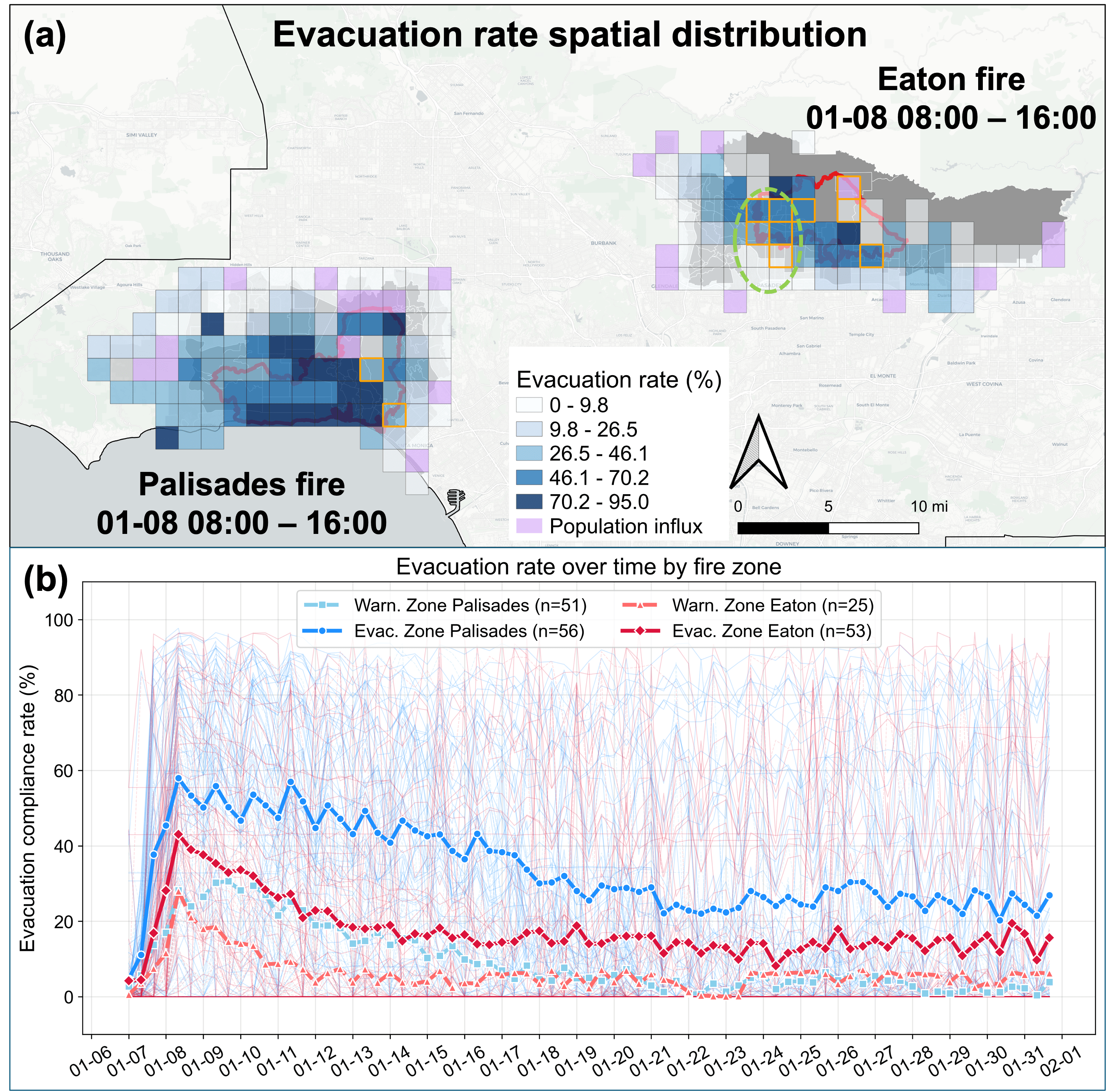}
  \caption{Spatiotemporal patterns of evacuation compliance rates during wildfire events. \textit{Notes}: (a) Spatial distribution of evacuation compliance rates for the Eaton and Palisades fires at the time when evacuation orders reached their maximum spatial extent. Orange boundaries highlight Bing tiles classified as wildfire vulnerability hotspots ($DEDI=1$), where high damage coincided with low evacuation rates. Pink tiles indicate areas that experienced a significant population influx during the wildfire events ($P_{i}>20\%$). The green circle highlights Bing tiles in West Altadena, where evacuation orders were issued with delay around midnight and corresponded to the highest fatality count (please further refer to Figure \ref{fig: Altadena_WE_evac_rate}). (b) Temporal changes in evacuation compliance rates for each Bing tile across different zones, with daily averages highlighted.}
  \label{fig: Evacuation_compliance_rate}
\end{figure}

Figure \ref{fig: Evacuation_compliance_rate} presents the spatiotemporal distribution of the evacuation compliance rates for both the Eaton and Palisades fires. The selected time slice corresponds to the daytime period on the second day after the two wildfires occurred. By this time, the evacuation order had largely covered the evacuation zones of the Palisades Fire, while the Eaton Fire also encompassed the main zones near the wildfire, as shown in Figure \ref{fig: evac_order_distribution}. Figure \ref{fig: Evacuation_compliance_rate}(a) shows the spatial variation in compliance across Bing tiles. Evacuation compliance rates ranged from 0\% to 95.0\%, with the highest level concentrated in areas adjacent to the fire perimeter, indicating clear spatial heterogeneity in evacuation behavior. Peripheral zones displayed moderate compliance rates (46.1–70.2\%), while distant warning zones showed minimal compliance (0–9.8\%). Several Bing tiles in outer warning zones exhibited population influx, reflecting outward migration during the early wildfire stages. In the Palisades fire, substantial displacement was observed in communities such as Malibu along State Route 1, Calabasas to the northwest, and Santa Monica to the southeast. The Eaton fire showed a similar spatial pattern, with high compliance observed in Pasadena and Sierra Madre, located west of the fire perimeter along U.S. Highway 210.

To examine the relationship between spatial proximity and evacuation compliance, Figure \ref{fig: corr_er_distance} plots evacuation compliance rates against the distance to the fire perimeter centroid. The result shows a moderate negative correlation ($r = -0.477$, $p < 0.001$), indicating that compliance rates decline as distance from the fire increases. This result supports the hypothesis that physical proximity to the fire is a key factor influencing evacuation behavior \cite{cova2024destination, sun2024social, wu2022wildfire, zhao2022estimating}.

\begin{figure}[!ht]
  \centering
  \includegraphics[width=0.8\textwidth]{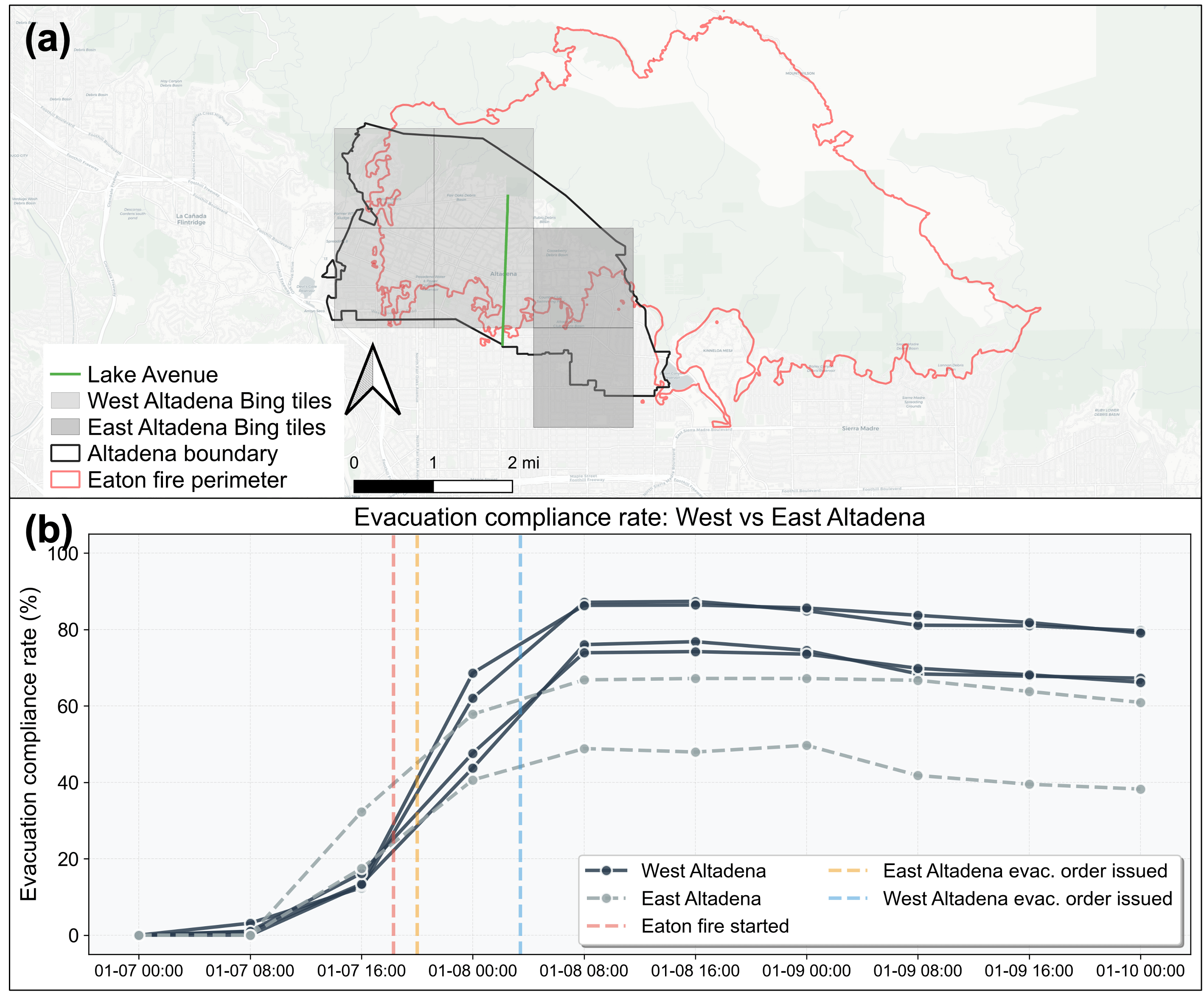}
  \caption{Spatial and temporal comparison of evacuation patterns in Altadena during the Eaton Fire. (a) Study area showing the Altadena boundary, divided into West and East Altadena Bing tiles relative to Lake Avenue, and the Eaton Fire perimeter. (b) Evacuation compliance rates for West and East Altadena from January 7 to January 10, 2025. Vertical dashed lines indicate the timing of the Eaton Fire ignition and the issuance of evacuation orders for each subarea.}
  \label{fig: Altadena_WE_evac_rate}
\end{figure}

In Figure \ref{fig: Evacuation_compliance_rate}(a), orange boundaries show the Bing tiles identified by DEDI. These tiles experienced high structural damage but low evacuation rates and are classified as wildfire vulnerability hotspots. Nine such tiles were located around the Eaton fire, and two were found in the Palisades fire. The green circle highlights five Bing tiles in West Altadena, an unincorporated area that received delayed and overnight evacuation orders \citep{nbc2025eaton, washpost2025altadena, yahoo2025altadena}. Residents in this area failed to evacuate promptly, resulting in low compliance rates. However, these tiles still experienced severe damage due to their direct exposure to the Eaton fire. Consequently, fifteen of the seventeen fatalities during the Eaton Fire occurred in West Altadena, west of Lake Avenue \citep{washpost2025altadena}. Figure \ref{fig: Altadena_WE_evac_rate} illustrates this pattern. Panel (a) identifies the corresponding Bing tiles in West and East Altadena, and panel (b) shows the changes in evacuation compliance rates over time for these six tiles. The seven-hour gap between the East (20:00 on January 7) and West (03:25 on January 8) evacuation orders generated a clear temporal divergence in response, with West Altadena exhibiting a delayed increase in compliance. Evacuation rates in the west began to rise only after the order was issued and peaked between 08:00 and 16:00 on January 8, one day after the fire started.

Figure \ref{fig: Evacuation_compliance_rate}(b) illustrates the temporal evolution of evacuation compliance rates from the onset of the wildfires to full containment. The analysis reveals distinct evacuation compliance patterns between two events. In the Palisades fire, evacuation zones exhibited higher initial compliance, reaching a peak of approximately 60\% within the first few days after orders were issued. Rates then declined and stabilized at approximately 30–35\% by the end of the observation period. In contrast, evacuation zones in the Eaton fire reached a lower peak of about 40\% and declined more gradually, stabilizing at around 15–20\%. Across both fires, compliance rates in warning zones remained consistently lower than in evacuation zones.

\begin{figure}[!ht]
  \centering
  \includegraphics[width=0.7\textwidth]{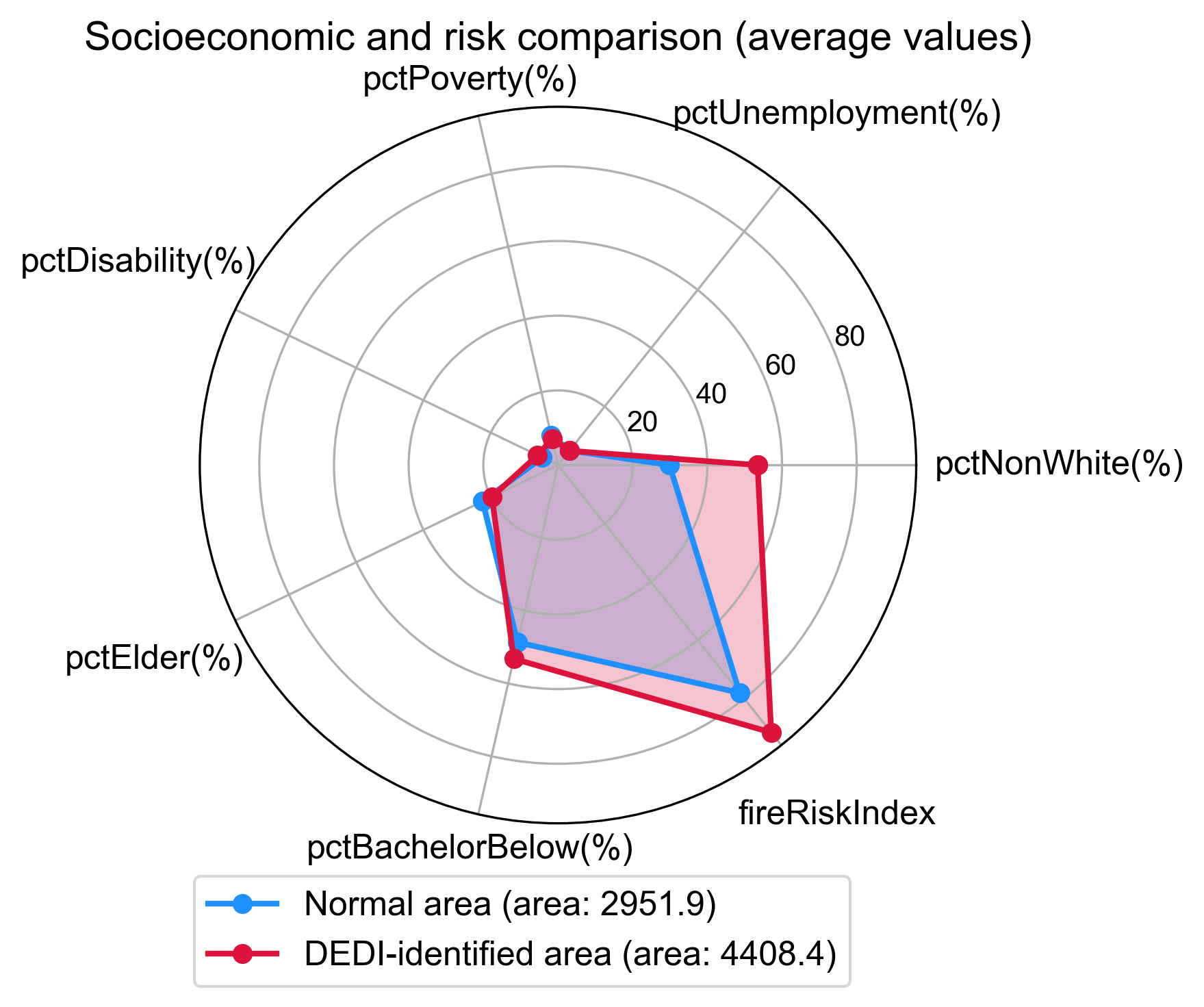}
  \caption{Radar chart of socioeconomic and wildfire risk indicators for normal areas and DEDI-identified (high damage low evacuation rate) areas. Variables include the percentages of non-White residents, unemployed individuals, people below the poverty line, people with disabilities, elders, residents without a bachelor's degree, and the fire risk index.}
  \label{fig: DEDI_radar_plot_comparison}
\end{figure}

Figure \ref{fig: DEDI_radar_plot_comparison} further compares the average socioeconomic and risk indicators of these DEDI-identified hotspots with other tiles. The results show that DEDI tiles generally exhibited higher average values, particularly for socioeconomic variables such as race and education, as well as for the fire risk index. This finding indicates that areas with high damage but low evacuation rates are often socially vulnerable communities, where low compliance highlights the need for targeted attention in evacuation planning.

\subsection{Evacuation departure time and delay}
The response time of household evacuation is often estimated using cumulative departure S-curves (e.g., the Rayleigh distribution) derived from empirical data collected during past disasters \citep{ozbay2012use, murray2013evacuation, woo2017reconstructing, zhao2022estimating}. In this study, we constructed cumulative evacuation response curves for each Bing tile within evacuation and warning zones to represent the proportion of evacuees departing at each time slice. As shown in Figure \ref{fig: evac_order_distribution}(b), we grouped Bing tiles into waves based on evacuation order issuance time to assess the influence of timing. The analysis period for the departure curves spans from wildfire onset through the following three days (January 8–11). Time is measured in hours since the fire started, enabling comparisons across zones regardless of order issuance timing.

\begin{figure}[!ht]
  \centering
  \includegraphics[width=\textwidth]{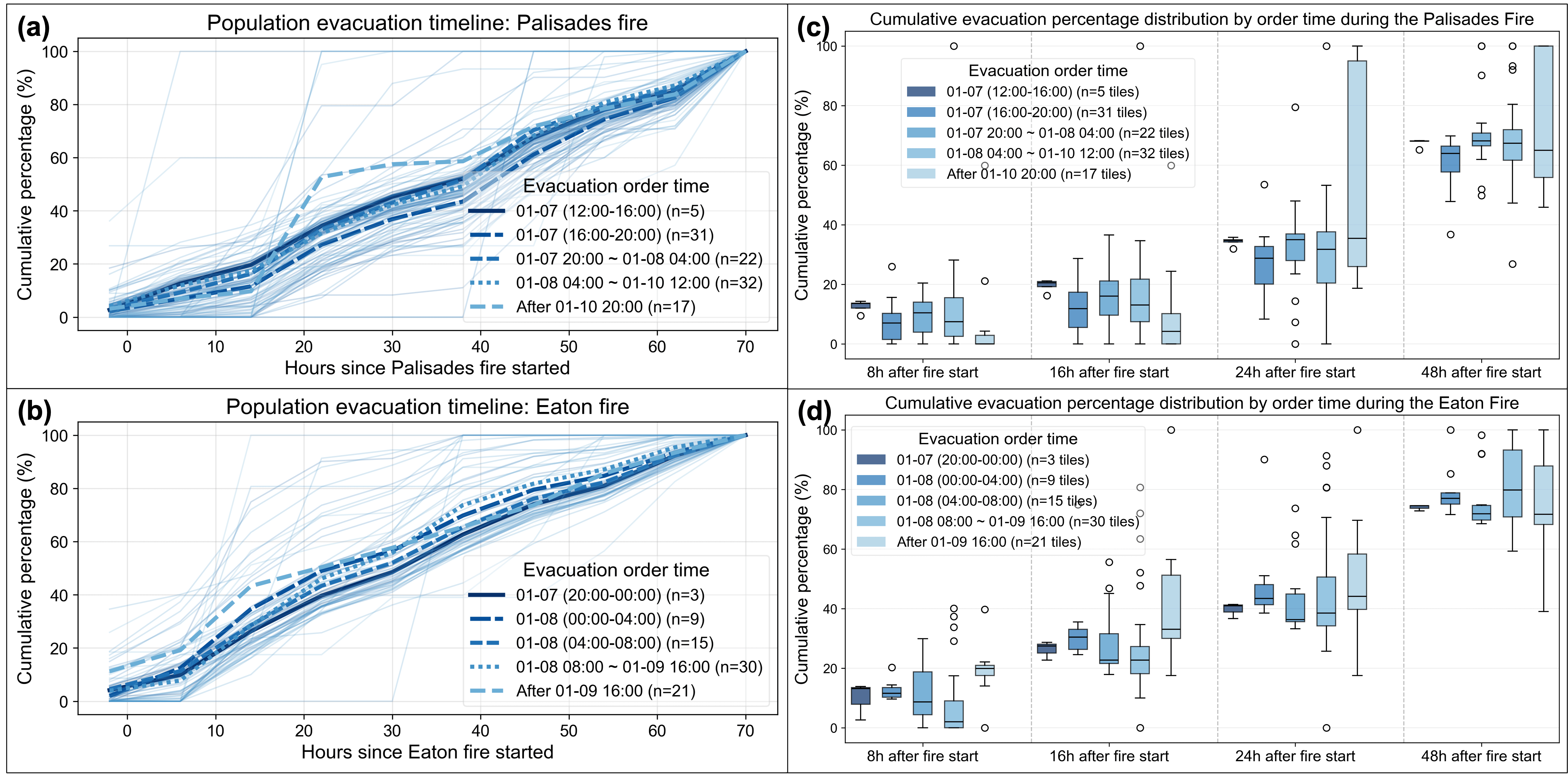}
  \caption{Cumulative evacuation departure curves grouped by evacuation order wave. \textit{Notes}: (a, b) Bing tiles are grouped into different waves based on the timing of their evacuation orders (see Figure \ref{fig: evac_order_distribution}(b)). Time is measured in hours since the fire started. Each curve represents the cumulative evacuation compliance rate of a single Bing tile, with bold lines showing the average trajectory for each wave. (c, d) Boxplots show the distribution of cumulative evacuation compliance rates for Bing tiles across different order waves at 8, 16, 24, and 48 hours after wildfire ignition.}
  \label{fig: S-curve}
\end{figure}

Figure \ref{fig: S-curve}(a) and (b) show cumulative evacuation curves for Bing tiles grouped by evacuation order wave. Earlier orders issuance corresponds to earlier departures and steeper cumulative curves. This pattern is clear in the Palisades fire. In contrast, most orders were issued around midnight on January 8 for the Eaton Fire, so the curves rose slowly before that time. Some tiles with orders issued after midnight on January 8 showed a share of self-evacuees who departed before official orders were issued \citep{zhao2022estimating}. Tiles that received the earliest orders display faster growth in departures. Both fires show marked increases in cumulative evacuation after the 16th hour of the Palisades Fire and the 8th hour of the Eaton Fire. These periods correspond to January 8 between 08:00 and 16:00, when all evacuation orders had been issued and movements took place during daytime hours.

Figure \ref{fig: S-curve}(c) and (d) show boxplots of cumulative evacuation percentages for Bing tiles at three time points: 8, 16, 24, and 48 hours after wildfire onset. In both fires, the cumulative evacuation rates reached about 40\% within 24 hours. Bing tiles that received earlier orders recorded high compliance. In the Eaton fire, tiles where orders were issued after midnight consistently maintained lower evacuation percentages. This pattern was especially evident in West Altadena, where late and nighttime orders reduced the timeliness of evacuations.

\begin{figure}[!ht]
  \centering
  \includegraphics[width=\textwidth]{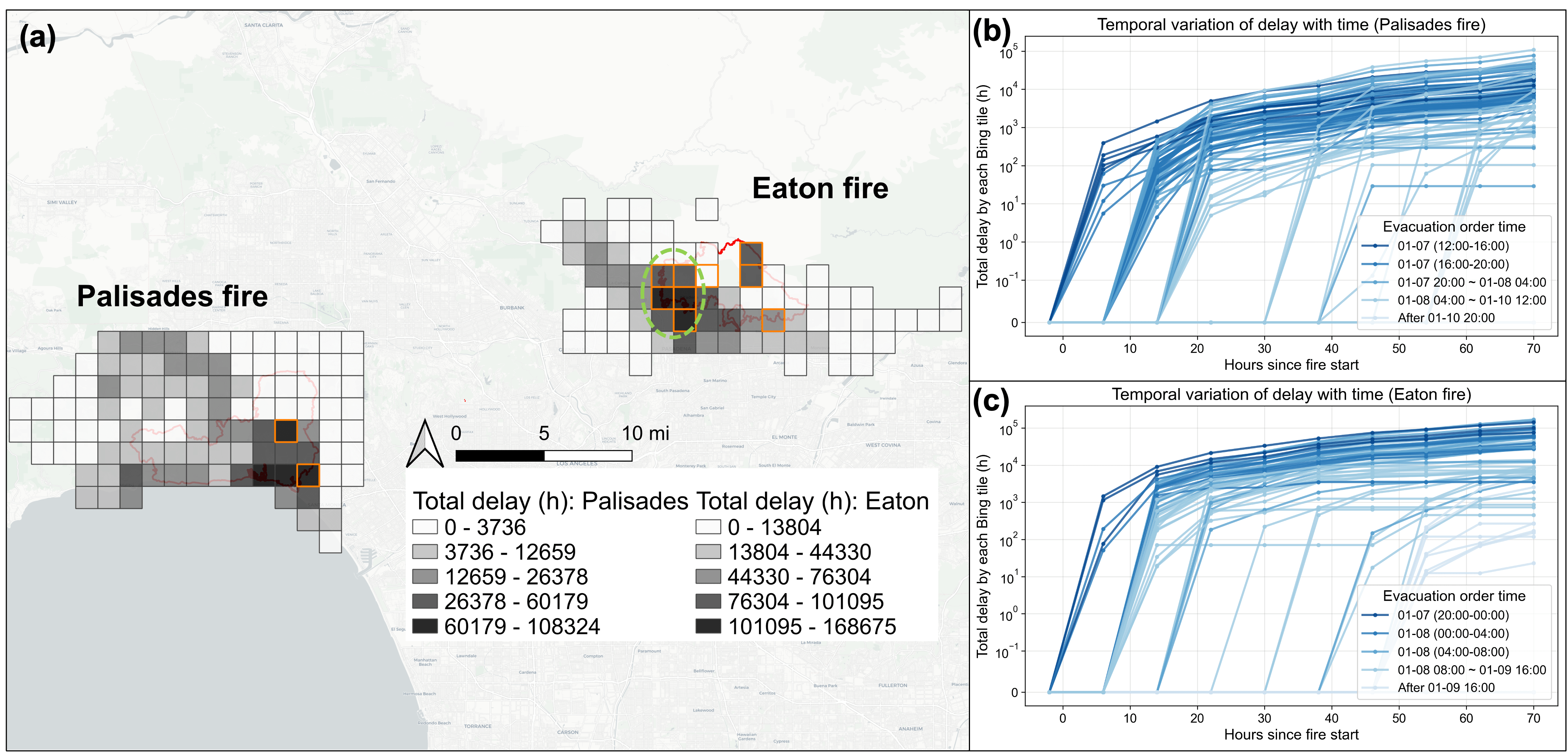}
  \caption{Cumulative evacuation departure delay: spatial distribution and temporal dynamics. \textit{Notes}: (a) Spatial distribution of total evacuation delays across Bing tiles. The green circle represents DEDI-identified tiles. (b, c) Temporal variation of departure delays for Bing tiles grouped by evacuation order issuance wave.}
  \label{fig: total_delay}
\end{figure}

Figure \ref{fig: total_delay}(a) illustrates the spatial distribution of total evacuation delays across Bing tiles. Tiles located near the fire perimeter show higher delays. In the Palisades fire, delays were concentrated in the southeastern areas bordering Pacific Palisades and Santa Monica. In the Eaton fire, delays were mainly observed in West Altadena and the northern part of La Cañada Flintridge. These areas lie within the wildfire–urban interface (WUI), are densely populated. Late evacuation orders further increased delays in these regions, as indicated by the green circles in Figure \ref{fig: total_delay}(a). We also observe that tiles marked with orange boundaries, identified by the DEDI metric, consistently exhibit high evacuation delays. This pattern suggests that socially vulnerable areas with lower evacuation rates tend to experience longer delays. Similar findings are reported in prior studies \citep{forrister2024analyzing, kuligowski2022modeling, sun2024social, xu2023predicting}.

Figure \ref{fig: total_delay}(b) and (c) display the temporal variation in total evacuation delays. In both fires, delay curves increase rapidly after orders are issued, but the timing differs. The Palisades Fire shows early acceleration following the January 7 orders at 12:00, 16:00, and 20:00. In contrast, the Eaton Fire delays rise primarily after midnight on January 8, reflecting later order issuance. These findings indicate that the timing of evacuation orders, combined with population density and location within the fire perimeter, strongly shaped the magnitude and distribution of evacuation delays.

\begin{figure}[!ht]
  \centering
  \includegraphics[width=\textwidth]{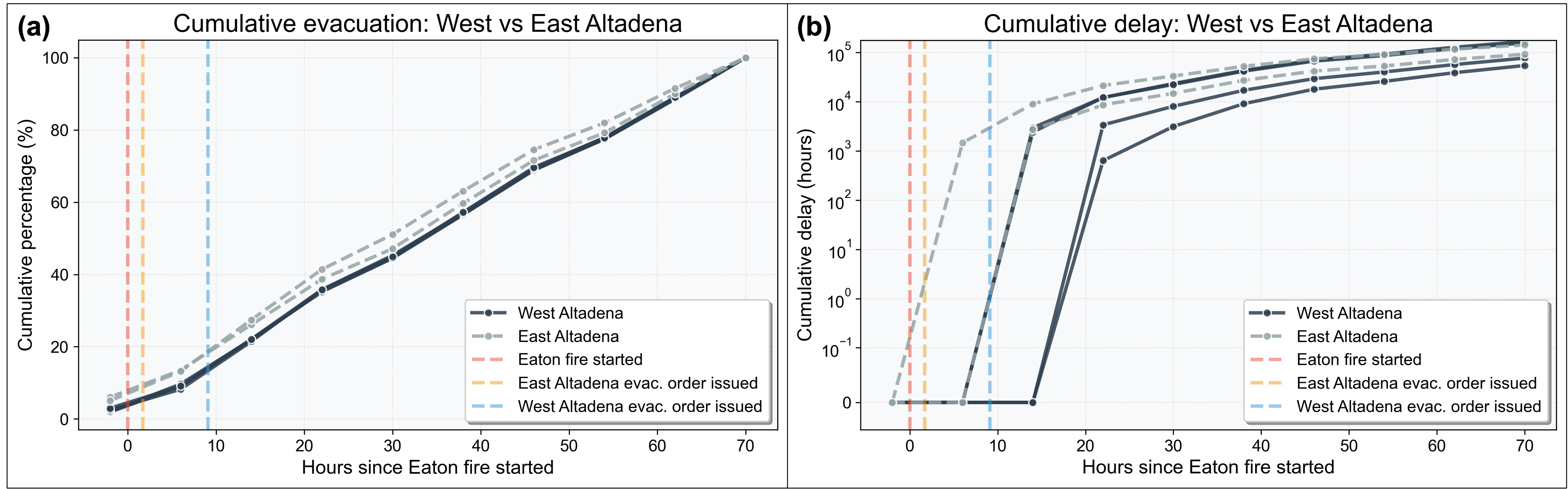}
  \caption{Cumulative evacuation and delay patterns in East and West Altadena during the Eaton Fire. \textit{Notes}: (a) Cumulative evacuation percentage over time. Please refer to Figure \ref{fig: Altadena_WE_evac_rate}(a) for the spatial distribution of Bing tiles. (b) Cumulative evacuation delay, showing the effect of the nine-hour difference in evacuation order issuance between the two subareas. Vertical dashed lines indicate the fire ignition and the issuance times of evacuation orders for each area.}
  \label{fig: Altadena_WE_CumEvac_Delay}
\end{figure}

To capture temporal differences in evacuation behavior between East and West Altadena, Figure \ref{fig: Altadena_WE_CumEvac_Delay} presents the cumulative evacuation and delay curves. East Altadena received its evacuation order nine hours earlier than West Altadena, and this difference is evident in the curves. The results reveal a distinct two-phase pattern: residents in East Altadena began evacuating soon after the initial order, whereas those in West Altadena showed a delayed rise in both cumulative departures and total delay. Some movement before the 3:25 a.m. order in West Altadena may reflect self-evacuation prompted by nearby fire cues.

\subsection{Evacuation movement spatial pattern}
\begin{figure}[!ht]
  \centering
  \includegraphics[width=\textwidth]{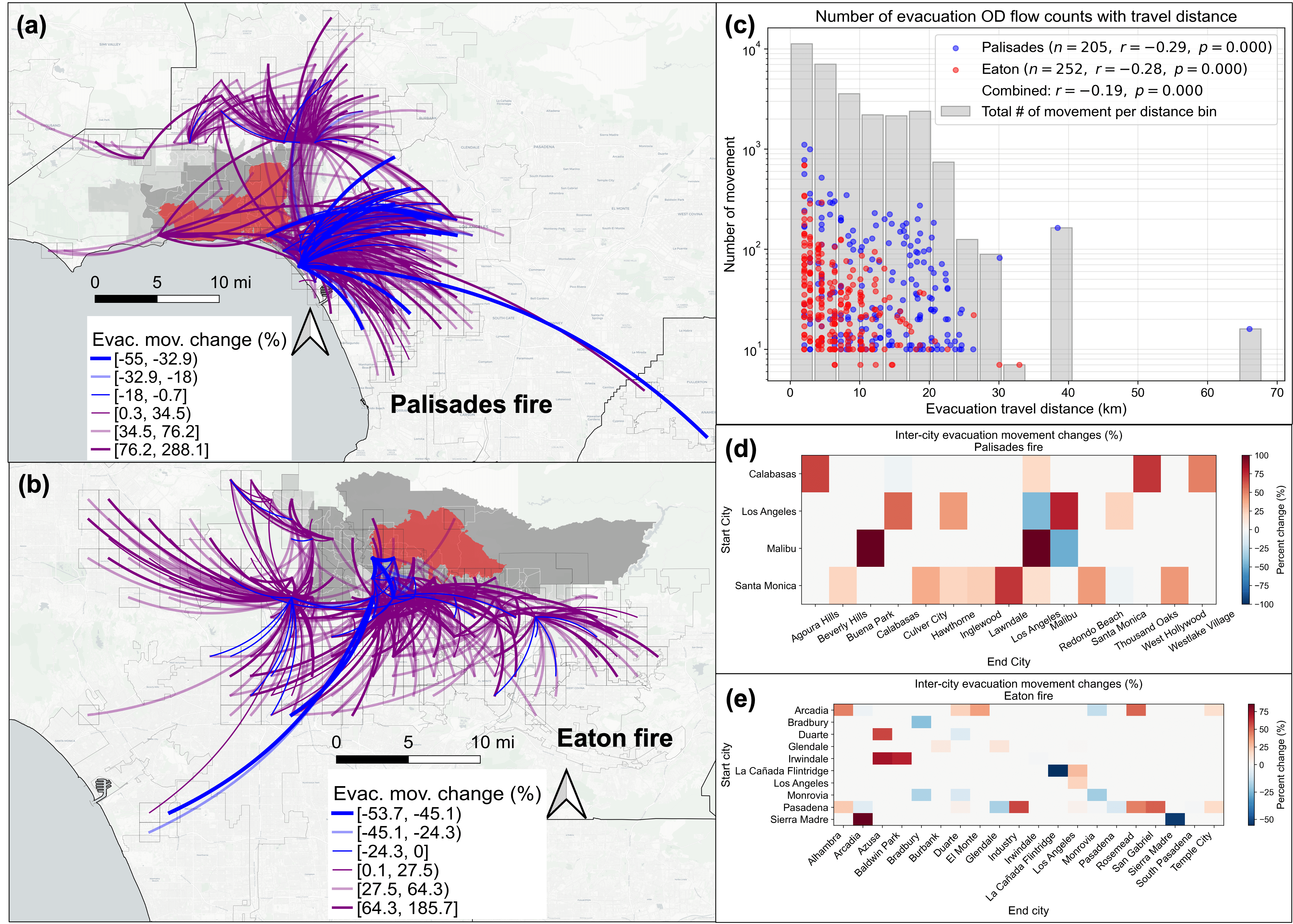}
  \caption{Spatial patterns of evacuation movement during wildfire events. \textit{Notes}: (a, b) Percent change in evacuation flows for the Palisades and Eaton fire. (c) Histogram and scatter plot of travel distance and flow counts. (d, e) Percent change in the number of evacuation movements between OD city pairs relative to baseline conditions.}
  \label{fig: mov_spatial_vis}
\end{figure}

Following the method outlined in Section \ref{section: evac_mov}, evacuation movements occurring within one week after the evacuation order were identified and mapped to illustrate their spatial distribution. Figure \ref{fig: mov_spatial_vis}(a) and (b) display the spatial distribution of percent changes in evacuation movement, and  Figure \ref{fig: mov_spatial_vis}(d) and (e) aggregate city-level OD percent changes. These patterns reveal distinct evacuation behaviors and highlight geographic disparities in population displacement between the two fire events. Both wildfires triggered substantial increases in evacuation movements relative to baseline conditions, with the Palisades fire generating a 12\% increase and the Eaton fire a 9\% increase. The Palisades fire (Figure \ref{fig: mov_spatial_vis}(a)) showed clear directional flows, with movements oriented mainly eastward and southward from the affected areas. The flow visualization highlights high-intensity evacuation corridors (dark purple, representing 76.2–288.1\% increases) extending from the fire perimeter toward east Los Angeles communities. The average evacuation distance increased sharply from 7.6 km at baseline to 9.3 km during the fire (22.3\% increase). This extended evacuation distance suggests that residents sought refuge farther away, likely due to the fire's proximity to densely populated coastal zones and limited nearby safe tiles. By contrast, the Eaton fire (Figure \ref{fig: mov_spatial_vis}(b)) displayed a more radial evacuation pattern with flows dispersing in multiple directions from the fire zone. The movements varied considerably in intensity, with some corridors experiencing extreme increases (64.3–185.7\%) while others showed more moderate changes. Although the overall increase in movement volume was smaller than that of the Palisades fire, the Eaton evacuation exhibited greater spatial complexity, with flows spread across multiple neighboring communities. The average evacuation distance increased from 4.4 km to 4.6 km (4.17\% increase), showing that evacuees generally found shelter closer to the fire zone.

Figure \ref{fig: mov_spatial_vis}(c) displays scatterplots and histograms illustrating the relationship between the number of movements and travel distance for the Eaton and Palisades fires. The number of movements decreases as the distance increases. This pattern indicates that most evacuation movements were short \citep{cova2024destination, wong2023understanding}. The Facebook movement dataset captures two notable peaks at 0–5 km and 15–20 km (see Figure \ref{fig: mov_long_short_trips_distribution}(a)). These peaks suggest distinct local and regional movement patterns. Figure \ref{fig: mov_spatial_vis}(d) and (e) illustrate inter-city evacuation movement percentage changes. Substantial evacuation flows occurred from Malibu to multiple destinations during the Palisades fire. Movements toward Los Angeles and Santa Monica were extreme and exceeded 100\%. The heatmap shows that coastal communities such as Santa Monica served as both origin and destination. This pattern reflects complex bidirectional movement dynamics. Pasadena and Sierra Madre served as the primary origins during the Eaton fire and generated significant outflows to nearby areas. Cities such as Arcadia and Monrovia experienced both inflows and outflows. This dual pattern suggests their role as evacuation origins and temporary refuge locations. These spatial patterns underscore the critical need for regional coordination in wildfire evacuation planning. Evacuation movements extend beyond the directly affected communities and create cascading effects across the broader metropolitan area. The differences between the two fires further illustrate how fire location, urban geography, and the structure of evacuation routes shape population displacement during wildfire events.

\subsection{Movement destination and distance}
Table \ref{tab: mov_des_dis} summarizes the proportion of evacuation destinations by land use type and the corresponding average evacuation distance using the method introduced in section \ref{section: evac_mov}. The land use classification of destination Bing tiles follows the approach of \cite{cova2024destination} as shown in Table \ref{tab: landClass}. The results reveal consistent patterns of destination choices and mobility behavior across the two wildfire events. Residential areas were the dominant destination type in both fires and accounting for 75.4\% and 73.8\% of movements during the Eaton and Palisades fires, respectively. This pattern underscores the fundamental role of social networks and residential resources in providing temporary shelter during wildfire evacuations. These resources include homes of friends and relatives as well as second homes \citep{toledo2018analysis, golshani2019analysis}.

\begin{table}[!ht]
\centering
\caption{Destination composition and evacuation travel distance by land use type.}
\label{tab: mov_des_dis}
\resizebox{\textwidth}{!}{%
\begin{tabular}{@{}llllllll@{}}
\toprule
 &  & \multicolumn{3}{c}{Palisades} & \multicolumn{3}{c}{Eaton} \\ \cmidrule(l){3-8} 
 & Land use & \%Share & Dis. & Corr. & \%Share & Dis. & Corr. \\ \midrule
Residential area & Residential area & \textbf{75.4} & 10.1 (7.8) & $-$0.205* & \textbf{73.8} & 6.3 (4.2) & $-$0.427*** \\ \midrule
Public facilities & Hotel/motel & 3.2 & \textbf{20.4 (9.2)} & 0.357** & 2.8 & \textbf{9.4 (5.5)} & 0.167* \\
 & Public facilities-others & 6.3 & 11.2 (7.8) & $-$0.106 & 6.9 & 6.8 (6.4) & $-$0.213* \\ \midrule
Commercial area & Shopping/store & 5.0 & 10.9 (7.5) & $-$0.072 & 4.9 & 7.1 (5.1) & $-$0.196 \\
 & Industry/working & 4.6 & 12.6 (8.7) & 0.032 & 6.2 & 9.1 (5.8) & 0.046 \\
 & Service & 4.6 & 14.9 (9.4) & $-$0.248 & 4.2 & 7.6 (5.8) & $-$0.124 \\ \midrule
Other types & Other types & 0.9 & 8.6 (4.2) & $-$0.198 & 1.2 & 9.2 (4.3) & 0.150 \\ \bottomrule
\end{tabular}%
}
\begin{flushleft}
\justifying
\footnotesize \textit{Notes}: (1) Detailed land use classifications are provided in Table \ref{tab: landClass}. (2) The ``Dis.'' column reports the mean evacuation travel distance (km), with standard deviation shown in parentheses. (3) Corr. indicates the Pearson correlation between the destination share and the corresponding travel distance across Bing tiles within each fire zone. (3) ***: $<$ 0.01; **: $<$ 0.05; *: $<$ 0.1.
\end{flushleft}
\end{table}

\begin{figure}[!ht]
  \centering
  \includegraphics[width=\textwidth]{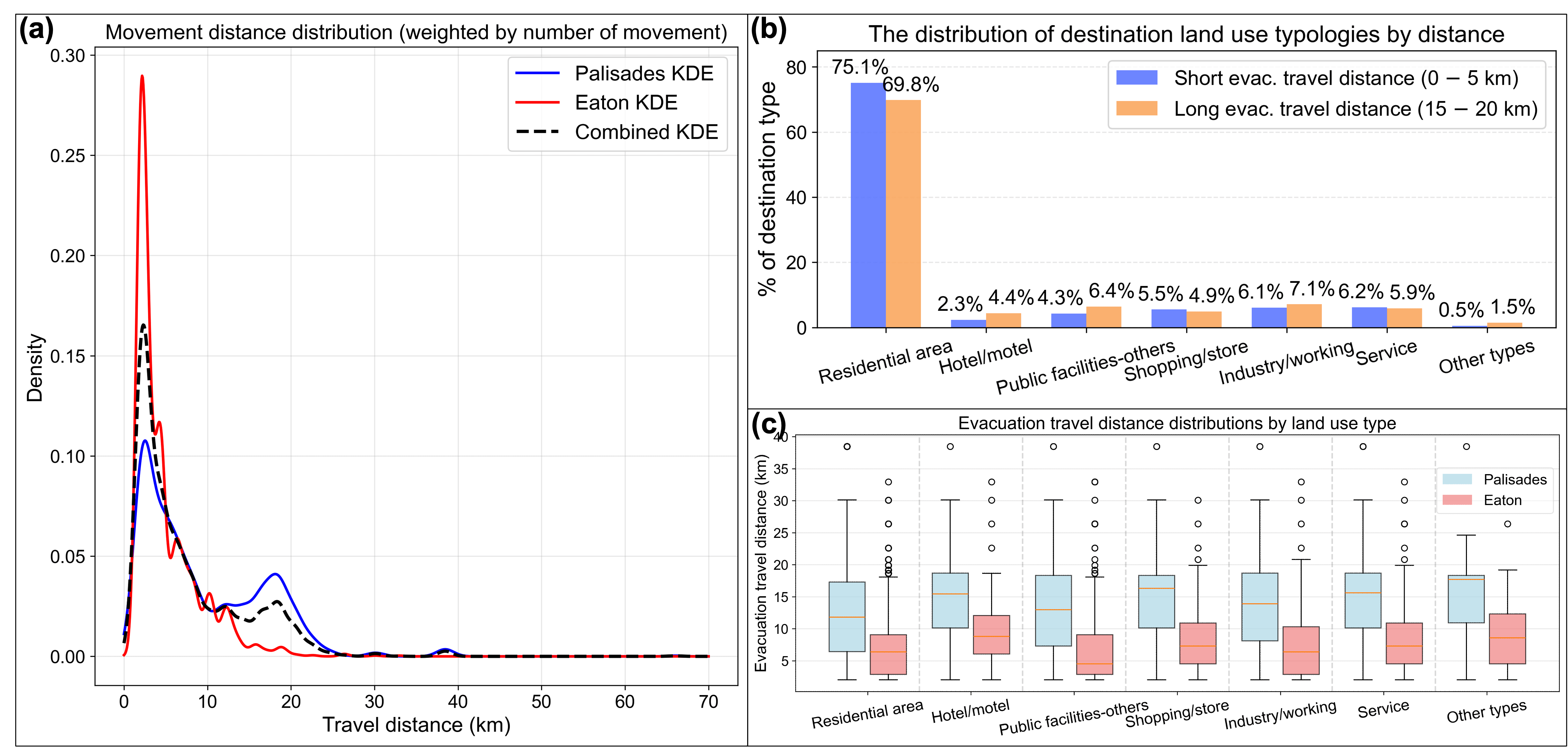}
  \caption{Travel distance and destination characteristics of wildfire evacuation. \textit{Notes}: (a) Kernel density distributions of evacuation travel distance for the Palisades and Eaton fires, weighted by the number of movements. Both distributions show strong peaks at short distance around 0–5 km and secondary peaks around 15–20 km. (b) Distribution of destination land-use typologies by travel distance. (c) Boxplots of evacuation travel distance by land-use type for the two fires.}
  \label{fig: mov_long_short_trips_distribution}
\end{figure}

In addition, a statistically significant negative correlation was found between the share of residential destinations and evacuation travel distance ($r = -0.427$, $p < 0.001$ for Eaton; $r = -0.205$, $p < 0.1$ for Palisades). Evacuees primarily sought refuge in nearby areas when possible. This tendency to minimize travel distance when choosing familiar or socially connected destinations supports the findings by \cite{cova2024destination}. In addition, the hotel/motel category showed a significant positive correlation with evacuation distance ($r = 0.357$, $p < 0.01$ for Eaton; $r = 0.167$, $p < 0.1$ for Palisades). This pattern confirms the association between longer-distance evacuation and formal public accommodations. Figure \ref{fig: mov_long_short_trips_distribution}(c) illustrates the distribution of evacuation travel distance by land use type through box plots that highlight variation within and across destination categories.

Public facilities and commercial areas accounted for comparable shares as secondary evacuation destinations. Public facilities constituted 9.7\% for the Eaton fire and 9.5\% for the Palisades fire. These locations include schools, hospitals, and government properties. They played a critical role in emergency management infrastructure by offering accessible shelter to evacuees without residential alternatives \citep{toledo2018analysis, beyki2023evacuation}. Commercial areas represented approximately 14–15\% of evacuation destinations in both fires. Industrial areas, working areas, and service establishments within this category exhibited consistently longer evacuation distance compared to shopping and retail locations.

However, we observed a relatively low proportion of evacuees selecting hotels or motels as evacuation destinations. This finding contrasts with \cite{cova2024destination}, who analyzed GPS data from the Kincade fire in Sonoma County, California, and identified hotels and motels as the primary type of public-facility destination after residential areas. One possible explanation relates to differences in population density and residential availability across regions. Unlike Sonoma County, which is sparsely populated, the areas affected by the Palisades fire (including Pacific Palisades, Malibu, and Santa Monica) are densely populated and wealthy. Communities near the Eaton fire show similar characteristics. These include Pasadena and La Cañada. Such communities offer more housing options nearby, such as the homes of friends, relatives, and coworkers. This access to informal accommodations may reduce reliance on hotels or motels during evacuations. Additionally, differences in data resolution may also contribute to this discrepancy. Our study utilized Facebook data aggregated at the Bing tile level to estimate destination proportions based on the areal coverage of each land use type. This approach lacks individual-level specificity and may underestimate hotel and motel selections. Despite these limitations, hotels and motels had the longest mean evacuation distance among all land-use categories in our findings. This positive association between selecting formal public accommodations and longer evacuation distance aligns with findings reported by \cite{cova2024destination}.

Further, Figure \ref{fig: mov_long_short_trips_distribution} presents the distribution of destinations by evacuation distance. Figure \ref{fig: mov_long_short_trips_distribution}(a) shows the kernel density distribution of evacuation movements based on all evacuation movements (Figure \ref{fig: mov_spatial_vis}(a) and (b)) weighted by the number of movements. The results reveal two distinct peaks in the relationship between evacuation movement frequency and distance during the Palisades and Eaton fires. These peaks occur at approximately 5 km and 15 km. Figure \ref{fig: mov_long_short_trips_distribution}(b) categorizes evacuation movements into short-distance and long-distance groups based on these peaks and compares their destination distributions. Residential areas dominated both groups. However, residential areas accounted for a higher proportion of short-distance movements than long-distance movements. Hotels and motels represented nearly twice the proportion of long-distance destinations compared to short-distance destinations. These patterns confirm the findings in Table \ref{tab: mov_des_dis} and align with \cite{cova2024destination}. In contrast, short-distance evacuation movements showed higher proportions of commercial areas such as shopping centers. Long-distance movements were more frequently directed toward industrial areas, working areas, and public facilities. These patterns reflect the distinct characteristics of wildfire evacuation destination choices. Figure \ref{fig: mov_long_short_trips_distribution}(c) presents boxplots of evacuation travel distance by land use type for the two fires. Movements toward residential and service areas are generally shorter, whereas movements toward industrial or public facilities tend to span longer distance. These findings further validate the results presented in Table \ref{tab: mov_des_dis}.

\section{Discussion} \label{Discussion}
\subsection{Major findings and implications}
This study used freely accessible, high-resolution Facebook datasets to analyze population evacuation behavior and mobility patterns during the 2025 Palisades and Eaton wildfires. Specifically, we employed the Facebook dataset to derive evacuation-related metrics. These metrics include evacuation compliance rates, departure timing, and delay. We then combined evacuation rates with structure damage data to identify vulnerability hotspot tiles characterized by high damage rates but low evacuation rates. Additionally, we utilized the Facebook movement dataset to analyze evacuation OD movement flows, destination choices, and travel distance. We further examined the distribution of evacuation travel distance and their relationship with destination types.

This paper provides several key insights. First, the proposed framework (Figure \ref{fig: Analysis framework}) demonstrates that Facebook datasets can effectively support analyses of population evacuation and mobility during wildfire events. The open-source nature and high spatiotemporal resolution of these datasets underscore their value and applicability to wildfire research. Although the Facebook data are aggregated at Bing tile level (2.4 km $\times$ 2.4 km), this spatial scale is comparable to the typical TAZ used in Southern California \citep{SCAG2024ConnectSoCal}. Therefore, the dataset provides an effective way of capturing wildfire evacuation behavior. Our results further validate Facebook data's capability to reflect evacuation dynamics, extending the findings of \cite{jia2020patterns}. Moreover, the analytical methods developed in this study are transferable to other disaster contexts, providing useful insights across different crisis scenarios.

Second, the proposed set of evacuation-related metrics is generalizable beyond the Facebook platform and can be adapted to other sudden-onset disaster contexts, such as hurricanes \citep{jamal2023understanding, rashid2025understanding}, floods \citep{maas2019facebook}, earthquakes \citep{varol2024movement}, and pandemics \citep{galeazzi2021human}. The metrics are scalable across spatial units ranging from Bing tiles to metropolitan regions (by aggregation), provided that population mobility or presence data are available at a comparable temporal resolution. In principle, any anonymized dataset that records human presence or movement dynamics over time can be used to compute these metrics \citep{liu2025hurricane, sun2024social, wu2022wildfire, zhao2022estimating}. The data should contain sufficient temporal resolution to capture population changes before and after the event, spatial consistency with the affected areas, and adequate granularity to infer shifts in movement or distribution through analytical modeling. These conditions can be satisfied by various data sources, such as GPS \citep{cova2024destination, zhao2022estimating} and traffic detector records \citep{feng2021reconstructing, dixit2014evacuation, janfeshanaraghi2025traffic}, highlighting the potential of the proposed metrics for wider disaster monitoring and cross-hazard comparison. However, the open accessibility and fine spatiotemporal granularity of Facebook data \citep{maas2019facebook} make it particularly well suited for real-time evacuation assessment.

Third, daily population changes captured by the Facebook population dataset illustrated spatiotemporal patterns of population evacuation at the Bing tile level during the Palisades and Eaton fires (Figure \ref{fig: Evacuation_compliance_rate}). These patterns highlight spatial heterogeneity in evacuation behavior driven by distinct wildfire characteristics and geographic conditions. Building on evacuation rates, we defined the Damage-Evacuation Disparity Index (DEDI) metric and combined it with structure damage data to identify Bing tiles with high damage rates but low evacuation rates. We found that these tiles were spatially concentrated near wildfire perimeters and temporally associated with late evacuation orders and nighttime order issuance. This pattern is evident in west Altadena affected by the Eaton Fire \citep{nbc2025eaton, yahoo2025altadena}, which contained 6 of the 11 identified high-damage, low-evacuation tiles (Figure \ref{fig: Evacuation_compliance_rate}). Figure \ref{fig: DEDI_radar_plot_comparison} further shows that DEDI-identified tiles typically exhibit higher social vulnerability and fire risk indices. This indicates that these areas require targeted attention in evacuation planning.

Fourth, we used daily population changes to generate cumulative evacuation response curves from wildfire onset through the following three days. These curves captured the critical period of wildfire evacuation (Figure \ref{fig: S-curve}). We combined these curves with evacuation order timing to calculate delay metrics (Figure \ref{fig: total_delay}). Results showed that later evacuation order issuance led to slower evacuation responses and greater delays. DEDI-identified tiles consistently exhibited high delay values (Figure \ref{fig: total_delay}(a)). Different wildfire characteristics and evacuation order timings resulted in unique temporal patterns in the evacuation curves. However, our analysis demonstrated that Facebook population data effectively capture evacuation timing, enriching existing wildfire evacuation research \citep{zhao2022estimating}.

Fifth, the Facebook movement dataset provided detailed insights into mobility patterns during wildfires. We quantified evacuation movement related metrics, including OD flows, travel distance, and destination distributions. We also examined changes in the volume and average distance of different movement types relative to normal conditions. Spatial visualizations (Figure \ref{fig: mov_spatial_vis}(a) and (b)) show clear differences between wildfire and baseline periods. Results showed that evacuation movements during the Palisades and Eaton fires increased by approximately 13\% in average distance and 10\% in frequency compared to normal conditions. Figure \ref{fig: mov_spatial_vis}(c) captured two distinct peaks in evacuation travel distance. Analyses of aggregated OD flows at the inter-city level (Figure \ref{fig: mov_spatial_vis}(d) and (e)) revealed trends in short-term inter-city migration. These visualizations help identify corridors of high evacuation traffic, informing targeted traffic management strategies.

Last, we analyzed evacuation destinations based on land-use classification methods from \cite{cova2024destination}. Specifically, we inferred destination proportions using the area-weighted land-use distribution within Bing tiles. We found that residential areas overwhelmingly dominated as evacuation destinations in both wildfire events. Residential destination shares showed significantly negative correlations with evacuation travel distance. In contrast, hotel/motel destinations showed significant positive correlations with travel distance. These patterns are consistent with previous studies \citep{cova2024destination, golshani2019analysis, toledo2018analysis}. For example, \cite{cova2024destination} used GPS data to track individual evacuees and found that longer-distance evacuations were more likely to end at public facilities such as hotels, parks, or federal properties, thereby reducing the share of residential destinations in such cases. This pattern suggests that evacuees tend to select socially connected and nearby residential locations as destinations whenever possible. However, when longer-distance travel is necessary, they are more likely to choose formal accommodations such as hotels or motels as temporary destinations along the evacuation route. To examine this pattern in more detail, we classified evacuation movements into short-distance (0–5 km) and long-distance (5–15 km) categories based on the relationship between evacuation travel distance and frequency identified in Figure \ref{fig: mov_spatial_vis}(c) and Figure \ref{fig: mov_long_short_trips_distribution}(a). We then examined the destination land-use distribution for each category (Figure \ref{fig: mov_long_short_trips_distribution}(b)). Results showed that long-distance evacuations more frequently selected hotels/motels and other public facilities as destinations, while short-distance evacuations predominantly chose residential areas such as second homes or relatives' homes. These findings corroborate the correlation patterns shown in Table \ref{tab: mov_des_dis}. Additionally, due to spatial aggregation in Facebook datasets, our results inevitably underestimate the proportion of evacuees choosing hotels or motels as destinations (Table \ref{tab: mov_des_dis}). We also found that hotel or motel destinations involved the longest average evacuation distance, consistent with prior research \citep{wong2023understanding, cova2024destination}.

This study also provides two practical recommendations for policymakers to enhance disaster management. First, policymakers can apply our findings and use openly accessible Facebook datasets to identify areas with high evacuation compliance rates and significant evacuation traffic flows. This approach enables real-time tracking of daily changes without incurring additional costs, helping policymakers effectively formulate, quantify, and validate recovery plans or budget allocations \citep{park2024post}. Second, policymakers can proactively plan for large-scale evacuations by targeting high-evacuation-demand areas with improved infrastructure and services. Special attention should be given to socially vulnerable communities with high damage rates but low evacuation rates, as well as those facing high wildfire risk. Recognizing that large evacuation movements cannot be entirely prevented, policymakers should minimize traffic congestion during critical evacuation periods by providing services such as bus-based evacuations for underserved populations. Additionally, disaster preparedness efforts should include educating residents about evacuation routes to reduce disorderly evacuations during emergency conditions.

\subsection{Limitations and future work}
Despite the several insights, this paper still contains several limitations. First, our analysis relies on aggregated social media datasets collected at the 2.4 km $\times $ 2.4 km grid cell level \citep{maas2019facebook}. To protect user privacy, Bing tiles with fewer than 10 users were excluded from the dataset. This level of aggregation and truncation constrains the precision of population density estimates, evacuation compliance rates calculations, and mobility analyses. However, the spatial area of Bing tiles remains comparable to typical traffic analysis zones in Southern California. This ensures the validity of our analytical conclusions. Future studies can address this limitation by incorporating disaggregated datasets such as GPS-based mobile phone data \citep{zhao2022estimating} or point-of-interest visitation records \citep{yabe2025behaviour}.

Second, as shown in Figures \ref{fig: S-curve} and \ref{fig: total_delay}, the temporal aggregation of Facebook data limits the precision in estimating evacuees' departure times and delays compared with survey data \citep{forrister2024analyzing} or GPS data \citep{sun2024social}. The 8-hour temporal resolution tends to overestimate evacuation timing, as reflected by the step-like patterns observed in the evacuation and delay curves. Nevertheless, the Facebook dataset analysis captures DEDI-identified Bing tiles as having high damage but low evacuation rates and reveals distinct spatial patterns in departure-time curves and cumulative delays across different tiles.

Third, spatial aggregation in Facebook data requires estimating destination type proportions using area-weighted calculations. This method significantly increases the estimated proportion of residential areas and underestimates the number of evacuees selecting hotels or motels as evacuation destinations \citep{cova2024destination} due to the land use types in Los Angeles County. This spatial aggregation method assumes that movements within Bing tiles have zero distance, further underestimating the actual evacuation travel distance. Moreover, since our analysis is conducted at the Bing tile level, explainable variables can only be inferred from aggregated CBG data using area-weighted averaging. This spatial aggregation further constrains our ability to detect fine-grained or nonlinear behavioral patterns.

\section{Conclusion} \label{Conclusion}
This study uses Facebook datasets to analyze evacuation and mobility patterns during the 2025 Palisades and Eaton wildfires in Los Angeles. We developed a transferable analytical framework for Facebook data that derives a series of evacuation-related metrics from travel generation and distribution perspectives. These metrics include evacuation compliance rate, departure time, delay, evacuation movement OD flows, distance, and destination distribution. We further combined structure damage data with evacuation rates to define the DEDI. This index enabled us to identify social vulnerability hotspots and characterize the socioeconomic features of these communities. By analyzing these wildfire evacuation metrics, we identified distinct behavioral patterns shaped by fire context, spatial proximity, and land use characteristics. We found that late evacuation orders and nighttime evacuations resulted in very low evacuation rates, high delays, and casualties. This pattern was evident in the high fatality rates observed in West Altadena during the Eaton Fire. Furthermore, residents closer to the fire exhibited higher compliance and faster evacuation, while distant residents showed delayed responses. Inferred evacuation destination patterns revealed that evacuees overwhelmingly chose residential destinations, often selecting nearby homes of relatives or friends. However, those traveling longer distance were more likely to seek shelter in hotels or public facilities. These findings reinforce the importance of social networks, proximity, and land use in evacuation decision-making. Our results offer insights into localized planning strategies and highlight the value of Facebook datasets to inform adaptive and context-sensitive evacuation management.

\section*{CRediT authorship contribution statement}
\textbf{Shangkun Jiang}: Writing – review \& editing, Writing – original draft, Visualization, Validation, Software, Methodology, Investigation, Formal analysis, Data curation, Conceptualization. \textbf{Ruggiero Lovreglio}: Writing - review \& editing, Formal analysis, Methodology, Conceptualization. \textbf{Thomas J. Cova}: Writing - review \& editing, Methodology, Formal analysis, Conceptualization. \textbf{Sangung Park}: Writing - review \& editing, Formal analysis. \textbf{Susu Xu}: Writing - review \& editing, Formal analysis. \textbf{Xilei Zhao}: Writing - review \& editing, Supervision, Funding acquisition, Conceptualization, Methodology, Formal analysis.

\section*{Declaration of competing interest}
The authors declare that they have no known competing financial interests or personal relationships that could have appeared to influence the work reported in this paper.

\section*{Acknowledgments}
This work was performed under the following financial assistance award No. 60NANB24D245 from U.S. Department of Commerce, National Institute of Standards and Technology (NIST). This material is based upon work supported by the U.S. National Science Foundation (NSF) under awards No. 2338959, 2536600, 2536601, and 2536602. Any opinions, findings, conclusions, or recommendations expressed in this material are those of the authors and do not necessarily reflect the views of NIST or NSF. During the preparation of this work, the authors used ChatGPT in order to check grammar errors and improve the language. After using this tool, the authors reviewed and edited the content as needed and took full responsibility for the content of the publication.

\section*{Appendix A. Timeline of the wildfires}
\renewcommand{\thetable}{A\arabic{table}}
\setcounter{table}{0}

% \begin{table}[H]
% \centering
% \caption{Timeline of the wildfires: key dates and impact.}
% \label{tab: timelineofwildfire}
% \resizebox{\textwidth}{!}{%
% \begin{tabular}{@{}lllllll@{}}
% \toprule
%  & \multicolumn{1}{c}{Onset date} & \multicolumn{1}{c}{Evac. order issued} & \multicolumn{1}{c}{Containment date} & \multicolumn{1}{c}{Burned area (acres)} & \multicolumn{1}{c}{Structures destroyed} & \multicolumn{1}{c}{Fatality} \\ \midrule
% Palisades fire & 01-07, 10:30 & 01-07, 12:44 & 01-31, 08:22 & 23,707 & 7,854 & 12 \\
% Eaton fire & 01-07, 18:18 & 01-07, 19:22 & 01-31, 18:54 & 14,021 & 10,491 & 17 \\ \bottomrule
% \end{tabular}%
% }
% \end{table}

\begin{table}[H]
\centering
\caption{Timeline of the wildfires: key dates and impact.}
\label{tab: timelineofwildfire}
\resizebox{\textwidth}{!}{%
\begin{tabular}{@{}lccccccc@{}}
\toprule
 & Onset date & \begin{tabular}[c]{@{}c@{}}Evac. order \\ issued\end{tabular} & \begin{tabular}[c]{@{}c@{}}Max evac. \\ order extend\end{tabular} & \begin{tabular}[c]{@{}c@{}}Containment \\ date\end{tabular} & \begin{tabular}[c]{@{}c@{}}Burned area \\ (acres)\end{tabular} & \begin{tabular}[c]{@{}c@{}}Structures \\ destroyed\end{tabular} & Fatality \\ \midrule
Palisades fire & \begin{tabular}[c]{@{}c@{}}01-07, \\ 10:30\end{tabular} & \begin{tabular}[c]{@{}c@{}}01-07, \\ 12:00\end{tabular} & \begin{tabular}[c]{@{}c@{}}01-11, \\ 20:13\end{tabular} & \begin{tabular}[c]{@{}c@{}}01-31, \\ 08:22\end{tabular} & 23,707 & 7,854 & 12 \\
Eaton fire & \begin{tabular}[c]{@{}c@{}}01-07, \\ 18:18\end{tabular} & \begin{tabular}[c]{@{}c@{}}01-07, \\ 19:27\end{tabular} & \begin{tabular}[c]{@{}c@{}}01-09, \\ 20:09\end{tabular} & \begin{tabular}[c]{@{}c@{}}01-31, \\ 18:54\end{tabular} & 14,021 & 10,491 & 17 \\ \bottomrule
\end{tabular}%
}
\end{table}

\renewcommand{\thefigure}{A\arabic{figure}}
\setcounter{figure}{0}

\begin{figure}[H]
  \centering
  \includegraphics[width=0.8\textwidth]{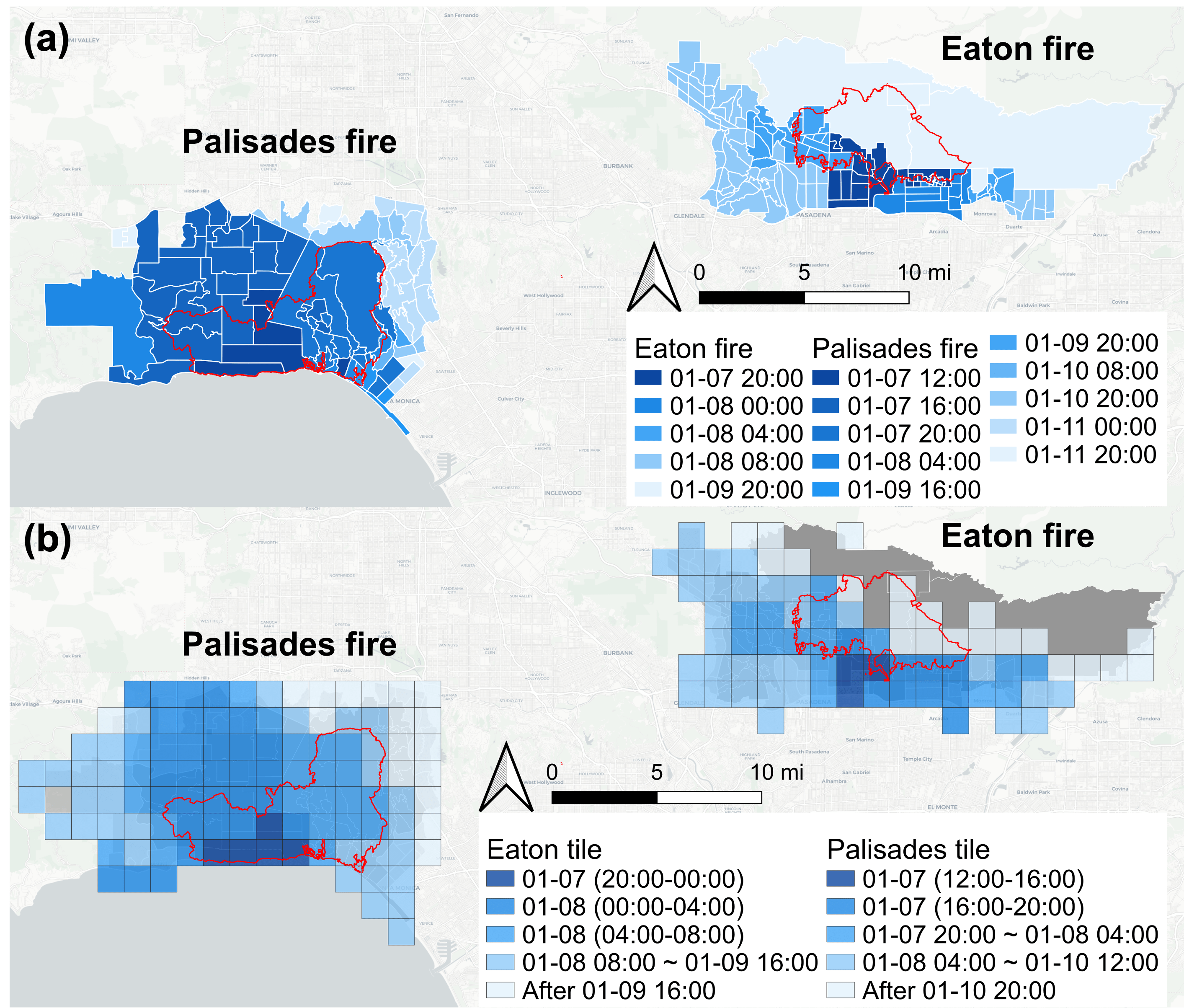}
  \caption{Temporal progression of evacuation order issuance and spatial distribution of affected areas. Source: \citep{calfire}. \textit{Notes}: (a) Temporal expansion of evacuation orders by zone; (b) corresponding Bing tile representation.}
  \label{fig: evac_order_distribution}
\end{figure}

\section*{Appendix B. Land-use type classification}
\renewcommand{\thetable}{B\arabic{table}}
\setcounter{table}{0}

\begin{table}[H]
\centering
\caption{Land-use type classification.}
\label{tab: landClass}
\resizebox{\textwidth}{!}{%
\begin{tabular}{@{}lll@{}}
\toprule
Primary   category & Secondary category & Detailed subtypes \\ \midrule
Residential area & - & - \\
Public facilities & Hotel/motel & - \\
 & Others & \begin{tabular}[c]{@{}l@{}}Federal property, School/University, Hospital, Church,  Airport,  Terminal, \\ Museum, Transmission facilities, Tower, Stadium, Government property, \\Library, Park, Beach, Harbor\end{tabular} \\
Commercial area & Shopping/store & - \\
 & Industry/working & Storage, Manufacture, Warehouse, Factory, Plant, Office, Processing \\
 & Service & \begin{tabular}[c]{@{}l@{}}Restaurant, Parking lot, Bank, Wash, Dealership,   Storage, Recreation,\\ Gas station, Repair, Gym/spa, Golf \end{tabular} \\
Other types & - & \begin{tabular}[c]{@{}l@{}} Vacant land, Mobile home, Agricultural land, Farm, Wetland, Cemetery\\ Desert, Waste, Miscellaneous \end{tabular} \\ \bottomrule
\end{tabular}%
}
\end{table}

\section*{Appendix C. Evacuation compliance rate vs distance analysis}
\renewcommand{\thefigure}{C\arabic{figure}}
\setcounter{figure}{0}

\begin{figure}[H]
  \centering
  \includegraphics[width=\textwidth]{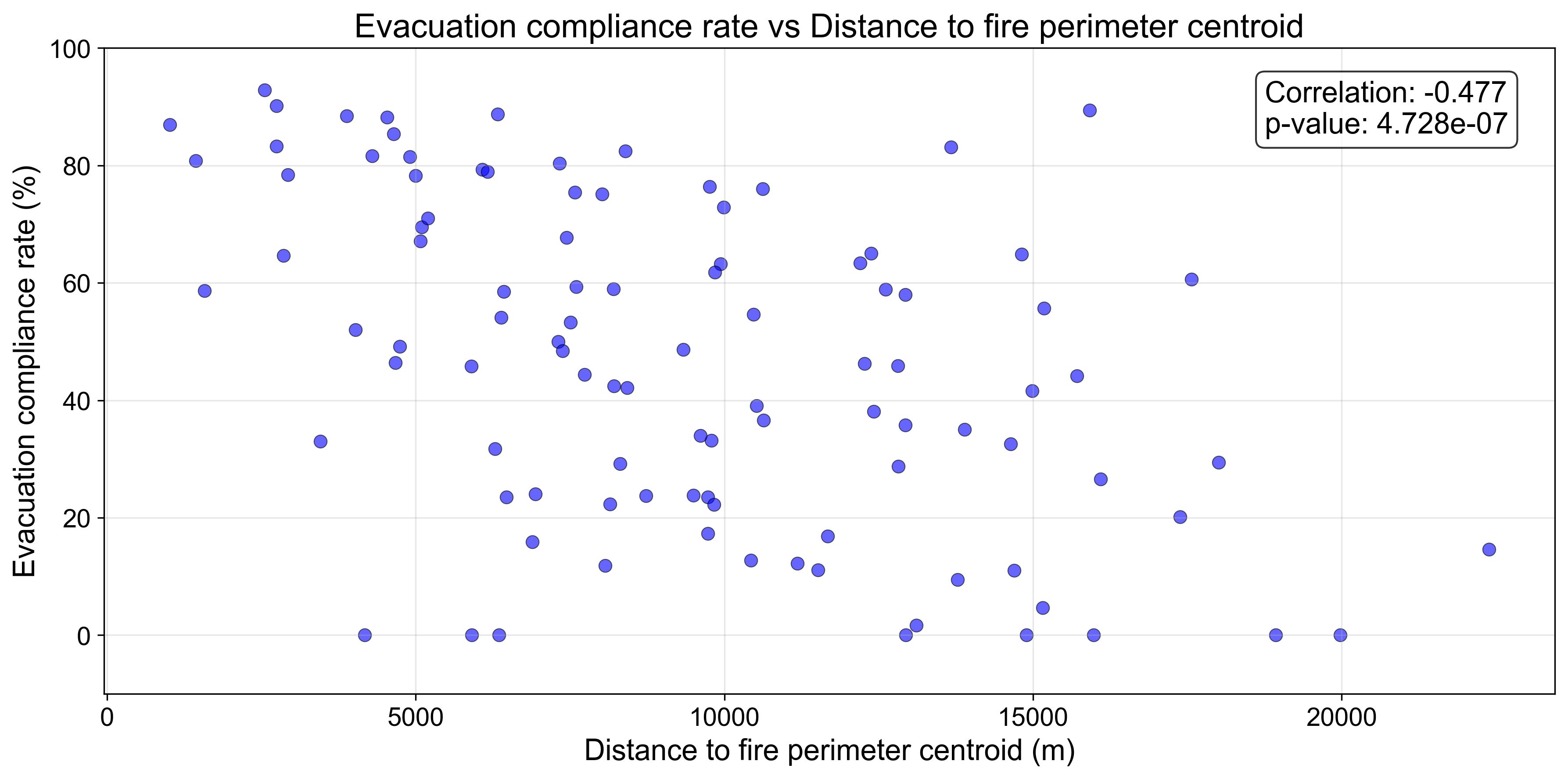}
  \caption{Relationship between evacuation compliance rate and distance to fire perimeter centroid.}
  \label{fig: corr_er_distance}
\end{figure}

\section*{Data availability}
The authors do not have permission to share data.

% \bibliographystyle{APA7} %elsarticle-num
% \bibliographystyle{elsarticle-num}

\bibliographystyle{elsarticle-harv}
\biboptions{semicolon,round,sort,authoryear} 
\bibliography{sample-base}

@inproceedings{maas2019facebook,
  title={Facebook Disaster Maps: Aggregate Insights for Crisis Response \& Recovery.},
  author={Maas, Paige and Iyer, Shankar and Gros, Andreas and Park, Wonhee and McGorman, Laura and Nayak, Chaya and Dow, P Alex},
  booktitle={KDD},
  volume={19},
  pages={3173},
  year={2019}
}

@misc{bingMapsTileSystem,
  title = {Bing Maps Tile System},
  year={2025},
  author={Joe Schwartz},
  howpublished = {\url{https://learn.microsoft.com/en-us/bingmaps/articles/bing-maps-tile-system}},
  note = {Accessed: 2025-03-07}
}

@article{achilleos2011inverse,
  title={The Inverse Distance Weighted interpolation method and error propagation mechanism--creating a DEM from an analogue topographical map},
  author={Achilleos, GA},
  journal={Journal of spatial Science},
  volume={56},
  number={2},
  pages={283--304},
  year={2011},
  publisher={Taylor \& Francis}
}

@article{jia2020patterns,
  title={Patterns of population displacement during mega-fires in California detected using Facebook Disaster Maps},
  author={Jia, Shenyue and Kim, Seung Hee and Nghiem, Son V and Doherty, Paul and Kafatos, Menas C},
  journal={Environmental Research Letters},
  volume={15},
  number={7},
  pages={074029},
  year={2020},
  publisher={IOP Publishing}
}

@article{jamal2023understanding,
  title={Understanding the loss in community resilience due to hurricanes using Facebook Data},
  author={Jamal, Tasnuba Binte and Hasan, Samiul},
  journal={International journal of disaster risk reduction},
  volume={97},
  pages={104036},
  year={2023},
  publisher={Elsevier}
}

@article{varol2024movement,
  title={The movement pattern changes of population following a disaster: Example of the Aegean Sea earthquake of October 2020},
  author={Varol, Cigdem and Hayrullahoglu, Gizem and Soylemez, Emrah and Sat, Necibe Aydan and Varol, Elif and Ozcan, Nazl{\i} Tunar},
  journal={International Journal of Disaster Risk Reduction},
  volume={112},
  pages={104743},
  year={2024},
  publisher={Elsevier}
}

@article{rashid2025understanding,
  title={Understanding hurricane evacuation behavior from Facebook data},
  author={Rashid, Md Mobasshir and Tirtha, Sudipta Dey and Eluru, Naveen and Hasan, Samiul},
  journal={International Journal of Disaster Risk Reduction},
  volume={116},
  pages={105147},
  year={2025},
  publisher={Elsevier}
}

@article{zhao2022estimating,
  title={Estimating wildfire evacuation decision and departure timing using large-scale GPS data},
  author={Zhao, Xilei and Xu, Yiming and Lovreglio, Ruggiero and Kuligowski, Erica and Nilsson, Daniel and Cova, Thomas J and Wu, Alex and Yan, Xiang},
  journal={Transportation research part D: transport and environment},
  volume={107},
  pages={103277},
  year={2022},
  publisher={Elsevier}
}

@misc{Caramela2025,
  title = {Pacific Palisades Wildfire Officially Most Destructive in Los Angeles History},
  year={2025},
  author={Sammi Caramela},
  howpublished = {\url{https://www.vice.com/en/article/pacific-palisades-wildfire-officially-most-destructive-in-los-angeles-history/}},
  note = {Accessed: 2025-03-07}
}

@misc{berker2025,
  title = {Southern California wildfires fully contained after 24 days},
  year={2025},
  author={Merve Berker},
  howpublished = {\url{https://www.aa.com.tr/en/americas/southern-california-wildfires-fully-contained-after-24-days/3468561}},
  note = {Accessed: 2025-03-07}
}

@misc{calfire,
  author       = {{CAL FIRE}},
  title        = {{CAL FIRE} — California Department of Forestry and Fire Protection},
  year         = {2025},
  howpublished = {\url{https://www.fire.ca.gov/}},
  note         = {Accessed: 2025-09-13},
  url         = {https://www.fire.ca.gov/},
  urldate      = {2025-09-13}
}

@article{duan2024identifying,
  title={Identifying counter-urbanisation using Facebook's user count data},
  author={Duan, Qianwen and Steele, Jessica and Cheng, Zhifeng and Cleary, Eimear and Ruktanonchai, Nick and Voepel, Hal and O'Riordan, Tim and Tatem, Andrew J and Sorichetta, Alessandro and Lai, Shengjie and others},
  journal={Habitat International},
  volume={150},
  pages={103113},
  year={2024},
  publisher={Elsevier}
}

@article{sun2024social,
  title={Social vulnerabilities and wildfire evacuations: A case study of the 2019 Kincade fire},
  author={Sun, Yuran and Forrister, Ana and Kuligowski, Erica D and Lovreglio, Ruggiero and Cova, Thomas J and Zhao, Xilei},
  journal={Safety Science},
  volume={176},
  pages={106557},
  year={2024},
  publisher={Elsevier}
}

@article{wu2022wildfire,
  title={Wildfire evacuation decision modeling using GPS data},
  author={Wu, Alex and Yan, Xiang and Kuligowski, Erica and Lovreglio, Ruggiero and Nilsson, Daniel and Cova, Thomas J and Xu, Yiming and Zhao, Xilei},
  journal={International Journal of Disaster Risk Reduction},
  volume={83},
  pages={103373},
  year={2022},
  publisher={Elsevier}
}

@article{yabe2020effects,
  title={Effects of income inequality on evacuation, reentry and segregation after disasters},
  author={Yabe, Takahiro and Ukkusuri, Satish V},
  journal={Transportation research part D: transport and environment},
  volume={82},
  pages={102260},
  year={2020},
  publisher={Elsevier}
}

@article{lindell2012protective,
  title={The protective action decision model: Theoretical modifications and additional evidence},
  author={Lindell, Michael K and Perry, Ronald W},
  journal={Risk Analysis: An International Journal},
  volume={32},
  number={4},
  pages={616--632},
  year={2012},
  publisher={Wiley Online Library}
}

@article{wong2023understanding,
  title={Understanding California wildfire evacuee behavior and joint choice making},
  author={Wong, Stephen D and Broader, Jacquelyn C and Walker, Joan L and Shaheen, Susan A},
  journal={Transportation},
  volume={50},
  number={4},
  pages={1165--1211},
  year={2023},
  publisher={Springer}
}

@article{cova2024destination,
  title={Destination unknown: Examining wildfire evacuee trips using GPS data},
  author={Cova, Thomas J and Sun, Yuran and Zhao, Xilei and Liu, Yepeng and Kuligowski, Erica D and Janfeshanaraghi, Nima and Lovreglio, Ruggiero},
  journal={Journal of Transport Geography},
  volume={117},
  pages={103863},
  year={2024},
  publisher={Elsevier}
}

@article{kuligowski2022modeling,
  title={Modeling evacuation decisions in the 2019 Kincade fire in California},
  author={Kuligowski, Erica D and Zhao, Xilei and Lovreglio, Ruggiero and Xu, Ningzhe and Yang, Kaitai and Westbury, Aaron and Nilsson, Daniel and Brown, Nancy},
  journal={Safety science},
  volume={146},
  pages={105541},
  year={2022},
  publisher={Elsevier}
}

@article{grajdura2022fast,
  title={Fast-moving dire wildfire evacuation simulation},
  author={Grajdura, Sarah and Borjigin, Sachraa and Niemeier, Deb},
  journal={Transportation research part D: transport and environment},
  volume={104},
  pages={103190},
  year={2022},
  publisher={Elsevier}
}

@article{radeloff2018rapid,
  title={Rapid growth of the US wildland-urban interface raises wildfire risk},
  author={Radeloff, Volker C and Helmers, David P and Kramer, H Anu and Mockrin, Miranda H and Alexandre, Patricia M and Bar-Massada, Avi and Butsic, Van and Hawbaker, Todd J and Martinuzzi, Sebasti{\'a}n and Syphard, Alexandra D and others},
  journal={Proceedings of the National Academy of Sciences},
  volume={115},
  number={13},
  pages={3314--3319},
  year={2018},
  publisher={National Academy of Sciences}
}

@article{ostertag2023investigating,
  title={Investigating spatiotemporal trends of large wildfires in California (1950--2020)},
  author={Ostertag, Sarah and Rice, Matthew and Qu, John J},
  journal={Advances in Cartography and GIScience of the ICA},
  volume={4},
  pages={16},
  year={2023},
  publisher={Copernicus Publications G{\"o}ttingen, Germany}
}

@article{xu2023predicting,
  title={Predicting and assessing wildfire evacuation decision-making using machine learning: Findings from the 2019 kincade fire},
  author={Xu, Ningzhe and Lovreglio, Ruggiero and Kuligowski, Erica D and Cova, Thomas J and Nilsson, Daniel and Zhao, Xilei},
  journal={Fire Technology},
  volume={59},
  number={2},
  pages={793--825},
  year={2023},
  publisher={Springer}
}

@article{grajdura2021awareness,
  title={Awareness, departure, and preparation time in no-notice wildfire evacuations},
  author={Grajdura, Sarah and Qian, Xiaodong and Niemeier, Deb},
  journal={Safety science},
  volume={139},
  pages={105258},
  year={2021},
  publisher={Elsevier}
}

@article{mclennan2019should,
  title={Should we leave now? Behavioral factors in evacuation under wildfire threat},
  author={McLennan, Jim and Ryan, Barbara and Bearman, Chris and Toh, Keith},
  journal={Fire technology},
  volume={55},
  pages={487--516},
  year={2019},
  publisher={Springer}
}

@article{wong2020can,
  title={Can sharing economy platforms increase social equity for vulnerable populations in disaster response and relief? A case study of the 2017 and 2018 California wildfires},
  author={Wong, Stephen D and Broader, Jacquelyn C and Shaheen, Susan A},
  journal={Transportation research interdisciplinary perspectives},
  volume={5},
  pages={100131},
  year={2020},
  publisher={Elsevier}
}

@article{toledo2018analysis,
  title={Analysis of evacuation behavior in a wildfire event},
  author={Toledo, Tomer and Marom, Ido and Grimberg, Einat and Bekhor, Shlomo},
  journal={International journal of disaster risk reduction},
  volume={31},
  pages={1366--1373},
  year={2018},
  publisher={Elsevier}
}

@article{christianson2019wildfire,
  title={Wildfire evacuation experiences of band members of whitefish lake first nation 459, Alberta, Canada},
  author={Christianson, Amy Cardinal and McGee, Tara K and Whitefish Lake First Nation 459},
  journal={Natural Hazards},
  volume={98},
  number={1},
  pages={9--29},
  year={2019},
  publisher={Springer}
}

@article{liu2025hurricane,
  title={Hurricane evacuation analysis with large-scale mobile device location data during hurricane Ian},
  author={Liu, Luyu and Zhang, Xiaojian and Jiang, Shangkun and Zhao, Xilei},
  journal={Transportation Research Part D: Transport and Environment},
  volume={139},
  pages={104559},
  year={2025},
  publisher={Elsevier}
}

@article{mccaffrey2018should,
  title={Should I stay or should I go now? Or should I wait and see? Influences on wildfire evacuation decisions},
  author={McCaffrey, Sarah and Wilson, Robyn and Konar, Avishek},
  journal={Risk analysis},
  volume={38},
  number={7},
  pages={1390--1404},
  year={2018},
  publisher={Wiley Online Library}
}

@article{bonaccorsi2020economic,
  title={Economic and social consequences of human mobility restrictions under COVID-19},
  author={Bonaccorsi, Giovanni and Pierri, Francesco and Cinelli, Matteo and Flori, Andrea and Galeazzi, Alessandro and Porcelli, Francesco and Schmidt, Ana Lucia and Valensise, Carlo Michele and Scala, Antonio and Quattrociocchi, Walter and others},
  journal={Proceedings of the national academy of sciences},
  volume={117},
  number={27},
  pages={15530--15535},
  year={2020},
  publisher={National Academy of Sciences}
}

@article{galeazzi2021human,
  title={Human mobility in response to COVID-19 in France, Italy and UK},
  author={Galeazzi, Alessandro and Cinelli, Matteo and Bonaccorsi, Giovanni and Pierri, Francesco and Schmidt, Ana Lucia and Scala, Antonio and Pammolli, Fabio and Quattrociocchi, Walter},
  journal={Scientific reports},
  volume={11},
  number={1},
  pages={13141},
  year={2021},
  publisher={Nature Publishing Group UK London}
}

@article{guan2025using,
  title={Using Multiple Biased Data Sets to Recover Missing Trips with a Behaviorally Informed Model},
  author={Guan, Xiangyang and Huang, Shuai and Chen, Cynthia},
  journal={Transportation Science},
  year={2025},
  publisher={INFORMS}
}

@article{forrister2024analyzing,
  title={Analyzing risk perception, evacuation decision and delay time: a case study of the 2021 Marshall Fire in Colorado},
  author={Forrister, Ana and Kuligowski, Erica D and Sun, Yuran and Yan, Xiang and Lovreglio, Ruggiero and Cova, Thomas J and Zhao, Xilei},
  journal={Travel behaviour and society},
  volume={35},
  pages={100729},
  year={2024},
  publisher={Elsevier}
}

@article{chen2014traces,
  title={From traces to trajectories: How well can we guess activity locations from mobile phone traces?},
  author={Chen, Cynthia and Bian, Ling and Ma, Jingtao},
  journal={Transportation Research Part C: Emerging Technologies},
  volume={46},
  pages={326--337},
  year={2014},
  publisher={Elsevier}
}

@article{park2024post,
  title={Post-disaster recovery policy assessment of urban socio-physical systems},
  author={Park, Sangung and Yabe, Takahiro and Ukkusuri, Satish V},
  journal={Computers, Environment and Urban Systems},
  volume={114},
  pages={102184},
  year={2024},
  publisher={Elsevier}
}

@article{murray2013evacuation,
  title={Evacuation transportation modeling: An overview of research, development, and practice},
  author={Murray-Tuite, Pamela and Wolshon, Brian},
  journal={Transportation Research Part C: Emerging Technologies},
  volume={27},
  pages={25--45},
  year={2013},
  publisher={Elsevier}
}

@article{ozbay2012use,
  title={Use of regional transportation planning tool for modeling emergency evacuation: Case study of northern New Jersey},
  author={Ozbay, Kaan and Yazici, M Anil and Iyer, Shrisan and Li, Jian and Ozguven, Eren Erman and Carnegie, Jon A},
  journal={Transportation research record},
  volume={2312},
  number={1},
  pages={89--97},
  year={2012},
  publisher={SAGE Publications Sage CA: Los Angeles, CA}
}

@article{woo2017reconstructing,
  title={Reconstructing an emergency evacuation by ground and air the wildfire in Fort McMurray, Alberta, Canada},
  author={Woo, Matthew and Hui, Kathy Tin Ying and Ren, Kexin and Gan, Kai Ernn and Kim, Amy},
  journal={Transportation Research Record},
  volume={2604},
  number={1},
  pages={63--70},
  year={2017},
  publisher={SAGE Publications Sage CA: Los Angeles, CA}
}

@article{golshani2019analysis,
  title={Analysis of evacuation destination and departure time choices for no-notice emergency events},
  author={Golshani, Nima and Shabanpour, Ramin and Mohammadian, Abolfazl and Auld, Joshua and Ley, Hubert},
  journal={Transportmetrica A: transport science},
  volume={15},
  number={2},
  pages={896--914},
  year={2019},
  publisher={Taylor \& Francis}
}

@article{beyki2023evacuation,
  title={Evacuation simulation under threat of wildfire—an overview of research, development, and knowledge gaps},
  author={Beyki, Shahab Mohammad and Santiago, Aldina and La{\'\i}m, Lu{\'\i}s and Craveiro, H{\'e}lder D},
  journal={Applied Sciences},
  volume={13},
  number={17},
  pages={9587},
  year={2023},
  publisher={MDPI}
}

@article{yabe2025behaviour,
  title={Behaviour-based dependency networks between places shape urban economic resilience},
  author={Yabe, Takahiro and Garc{\'\i}a Bulle Bueno, Bernardo and Frank, Morgan R and Pentland, Alex and Moro, Esteban},
  journal={Nature human behaviour},
  volume={9},
  number={3},
  pages={496--506},
  year={2025},
  publisher={Nature Publishing Group UK London}
}

@article{jones2024global,
  title={Global rise in forest fire emissions linked to climate change in the extratropics},
  author={Jones, Matthew W and Veraverbeke, Sander and Andela, Niels and Doerr, Stefan H and Kolden, Crystal and Mataveli, Guilherme and Pettinari, M Lucrecia and Le Qu{\'e}r{\'e}, Corinne and Rosan, Thais M and van der Werf, Guido R and others},
  journal={Science},
  volume={386},
  number={6719},
  pages={eadl5889},
  year={2024},
  publisher={American Association for the Advancement of Science}
}

@article{bowman2023taming,
  title={Taming the flame, from local to global extreme wildfires},
  author={Bowman, David MJS and Sharples, Jason J},
  journal={Science},
  volume={381},
  number={6658},
  pages={616--619},
  year={2023},
  publisher={American Association for the Advancement of Science}
}

@article{brachman2020wayfinding,
  title={Wayfinding during a wildfire evacuation},
  author={Brachman, Micah L and Church, Richard and Adams, Benjamin and Bassett, Danielle},
  journal={Disaster Prevention and Management: An International Journal},
  volume={29},
  number={3},
  pages={249--265},
  year={2020},
  publisher={Emerald Publishing Limited}
}

@online{washpost2025altadena,
  author       = {Joyce Sohyun Lee and Joshua Partlow and others},
  title        = {What went wrong the night Altadena burned},
  year         = {2025},
  howpublished = {\url{https://www.washingtonpost.com/weather/interactive/2025/altadena-wildfire-destruction-eaton-fire/}},
  note         = {Accessed: 2025-10-02}
}

@online{nbc2025eaton,
  author       = {Erik Ortiz and Jon Schuppe},
  title        = {All 17 deaths in Eaton Fire were in a zone where evacuation orders took hours to arrive},
  year         = {2025},
  howpublished = {\url{https://www.nbcnews.com/news/us-news/eaton-fire-deaths-los-angeles-evacuation-orders-took-hours-rcna188729}},
  note         = {Accessed: 2025-10-02}
}

@online{yahoo2025altadena,
  author       = {Terry Castleman, Ian James},
  title        = {Western Altadena Got Evacuation Order — But It Took Hours},
  year         = {2025},
  howpublished = {\url{https://www.yahoo.com/news/western-altadena-got-evacuation-order-004500325.html}},
  note         = {Accessed: 2025-10-02}
}

@techreport{SCAG2024ConnectSoCal,
  title        = {Connect SoCal 2024: Demographics and Growth Forecast Technical Report},
  author       = {{Southern California Association of Governments}},
  year         = {2024},
  institution  = {Southern California Association of Governments},
  address      = {Los Angeles, CA},
  note         = {Adopted April 4, 2024},
  url          = {https://www.scag.ca.gov/sites/default/files/2024-05/23-2987-tr-demographics-growth-forecast-final-040424.pdf#page=3.11}
}

@article{liu2024association,
  title={Association between NO2 and human mobility: a two-year spatiotemporal study during the COVID-19 pandemic in Southeast Asia},
  author={Liu, Zhaoyin and Li, Yangyang and Law, Andrea and Tan, Jia Yu Karen and Chua, Wee Han and Zhu, Yihan and Feng, Chen-Chieh and Luo, Wei},
  journal={Annals of GIS},
  volume={30},
  number={4},
  pages={475--492},
  year={2024},
  publisher={Taylor \& Francis}
}

@article{lusseau2023disparities,
  title={Disparities in greenspace access during COVID-19 mobility restrictions},
  author={Lusseau, David and Baillie, Rosie},
  journal={Environmental Research},
  volume={225},
  pages={115551},
  year={2023},
  publisher={Elsevier}
}

@article{beria2021presence,
  title={Presence and mobility of the population during the first wave of Covid-19 outbreak and lockdown in Italy},
  author={Beria, Paolo and Lunkar, Vardhman},
  journal={Sustainable cities and society},
  volume={65},
  pages={102616},
  year={2021},
  publisher={Elsevier}
}

@article{rashid2025network,
  title={Network-Wide Evacuation Traffic Prediction in a Rapidly Intensifying Hurricane from Traffic Detectors and Facebook Movement Data: Deep-Learning Approach},
  author={Rashid, Md Mobasshir and Rahman, Rezaur and Hasan, Samiul},
  journal={Journal of Transportation Engineering, Part A: Systems},
  volume={151},
  number={1},
  pages={04024085},
  year={2025},
  publisher={American Society of Civil Engineers}
}

@article{dixit2014evacuation,
  title={Evacuation traffic dynamics},
  author={Dixit, Vinayak and Wolshon, Brian},
  journal={Transportation research part C: emerging technologies},
  volume={49},
  pages={114--125},
  year={2014},
  publisher={Elsevier}
}

@article{rohaert2023traffic,
  title={Traffic dynamics during the 2019 Kincade wildfire evacuation},
  author={Rohaert, Arthur and Kuligowski, Erica D and Ardinge, Adam and Wahlqvist, Jonathan and Gwynne, Steven MV and Kimball, Amanda and B{\'e}nichou, Noureddine and Ronchi, Enrico},
  journal={Transportation research part D: transport and environment},
  volume={116},
  pages={103610},
  year={2023},
  publisher={Elsevier}
}

@article{rohaert2023analysis,
  title={The analysis of traffic data of wildfire evacuation: the case study of the 2020 Glass Fire},
  author={Rohaert, Arthur and Janfeshanaraghi, Nima and Kuligowski, Erica and Ronchi, Enrico},
  journal={Fire safety journal},
  volume={141},
  pages={103909},
  year={2023},
  publisher={Elsevier}
}

@article{janfeshanaraghi2025traffic,
  title={Traffic Performance Indicators for Evacuation: The Case Study of the 2020 Silverado Wildfire},
  author={Janfeshanaraghi, Nima and Rohaert, Arthur and Ronchi, Enrico and B{\'e}nichou, Noureddine and Kuligowski, Erica D},
  journal={Fire Technology},
  pages={1--25},
  year={2025},
  publisher={Springer}
}

@article{naser2025vulnerability,
  title={Vulnerability of structures and infrastructure to wildfires: a perspective into assessment and mitigation strategies},
  author={Naser, MZ and Kodur, Venkatesh},
  journal={Natural Hazards},
  volume={121},
  number={8},
  pages={9995--10015},
  year={2025},
  publisher={Springer}
}

@article{paveglio2016evaluating,
  title={Evaluating the characteristics of social vulnerability to wildfire: Demographics, perceptions, and parcel characteristics},
  author={Paveglio, Travis B and Prato, Tony and Edgeley, Catrin and Nalle, Darek},
  journal={Environmental Management},
  volume={58},
  number={3},
  pages={534--548},
  year={2016},
  publisher={Springer}
}

@article{lambrou2023social,
  title={Social drivers of vulnerability to wildfire disasters: A review of the literature},
  author={Lambrou, Nicole and Kolden, Crystal and Loukaitou-Sideris, Anastasia and Anjum, Erica and Acey, Charisma},
  journal={Landscape and Urban Planning},
  volume={237},
  pages={104797},
  year={2023},
  publisher={Elsevier}
}

@article{kuligowski2021evacuation,
  title={Evacuation decision-making and behavior in wildfires: Past research, current challenges and a future research agenda},
  author={Kuligowski, Erica},
  journal={Fire Safety Journal},
  volume={120},
  pages={103129},
  year={2021},
  publisher={Elsevier}
}

@article{palaiologou2019social,
  title={Social vulnerability to large wildfires in the western USA},
  author={Palaiologou, Palaiologos and Ager, Alan A and Nielsen-Pincus, Max and Evers, Cody R and Day, Michelle A},
  journal={Landscape and urban planning},
  volume={189},
  pages={99--116},
  year={2019},
  publisher={Elsevier}
}

@article{kuligowski2013predicting,
  title={Predicting human behavior during fires},
  author={Kuligowski, Erica},
  journal={Fire technology},
  volume={49},
  number={1},
  pages={101--120},
  year={2013},
  publisher={Springer}
}

@article{raei2025data,
  title={Are the data good enough? Spatial and temporal modeling of evacuee behavior using GPS data in a small rural community},
  author={Raei, Bahareh and Kinateder, Max and Benichou, Noureddine and Gomaa, Islam and Wang, Xin},
  journal={International Journal of Disaster Risk Reduction},
  volume={116},
  pages={105054},
  year={2025},
  publisher={Elsevier}
}

@article{feng2022modeling,
  title={Modeling and analyzing the traffic flow during evacuation in Hurricane Irma (2017)},
  author={Feng, Kairui and Lin, Ning},
  journal={Transportation research part D: transport and environment},
  volume={110},
  pages={103412},
  year={2022},
  publisher={Elsevier}
}

@article{feng2021reconstructing,
  title={Reconstructing and analyzing the traffic flow during evacuation in Hurricane Irma (2017)},
  author={Feng, Kairui and Lin, Ning},
  journal={Transportation research part D: transport and environment},
  volume={94},
  pages={102788},
  year={2021},
  publisher={Elsevier}
}

@article{lovreglio2019modelling,
  title={A modelling framework for householder decision-making for wildfire emergencies},
  author={Lovreglio, Ruggiero and Kuligowski, Erica and Gwynne, Steve and Strahan, Ken},
  journal={International Journal of Disaster Risk Reduction},
  volume={41},
  pages={101274},
  year={2019},
  publisher={Elsevier}
}

@article{ronchi2019open,
  title={An open multi-physics framework for modelling wildland-urban interface fire evacuations},
  author={Ronchi, Enrico and Gwynne, Steven MV and Rein, Guillermo and Intini, Paolo and Wadhwani, Rahul},
  journal={Safety science},
  volume={118},
  pages={868--880},
  year={2019},
  publisher={Elsevier}
}

@article{whittaker2017experiences,
  title={Experiences of sheltering during the Black Saturday bushfires: Implications for policy and research},
  author={Whittaker, Joshua and Blanchi, Raphaele and Haynes, Katharine and Leonard, Justin and Opie, Kimberley},
  journal={International journal of disaster risk reduction},
  volume={23},
  pages={119--127},
  year={2017},
  publisher={Elsevier}
}

@article{ahmad2024evaluating,
  title={Evaluating driving behavior patterns during wildfire evacuations in wildland-urban interface zones using connected vehicles data},
  author={Ahmad, Salman and Ahmed, Hafiz Usman and Ali, Asad and Yang, Xinyi and Huang, Ying and Guo, Mingwei and Ren, Yihao and Lu, Pan},
  journal={Fire Safety Journal},
  volume={142},
  pages={104015},
  year={2024},
  publisher={Elsevier}
}

@article{mahmoud2024leveraging,
  title={Leveraging epidemic network models towards wildfire resilience},
  author={Mahmoud, Hussam},
  journal={Nature Computational Science},
  volume={4},
  number={4},
  pages={253--256},
  year={2024},
  publisher={Nature Publishing Group US New York}
}

@article{mahmoud2020barriers,
  title={Barriers to gauging built environment climate vulnerability},
  author={Mahmoud, Hussam},
  journal={Nature Climate Change},
  volume={10},
  number={6},
  pages={482--485},
  year={2020},
  publisher={Nature Publishing Group UK London}
}




% %% For numbered reference style

%   \textit{\LaTeX: a document preparation system},
%   Addison Wesley, Massachusetts,

% \end{thebibliography}
\end{document}